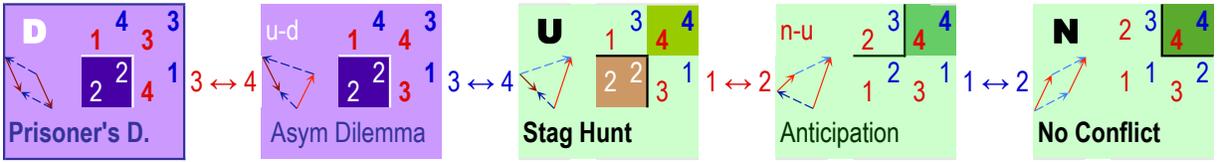

# Escaping Prisoner's Dilemmas:
## From Discord to Harmony
## in the Landscape of 2x2 Games

### Bryan Bruns
bryanbruns@bryanbruns.com
www.bryanbruns.com/2x2chart

Working Paper - June 8, 2012[1]

*Abstract: Changes in payoffs can transform Prisoner's Dilemma and other social dilemmas into harmonious win-win games. Using the Robinson-Goforth topology of 2x2 games, this paper analyzes how payoff swaps turn Prisoner's Dilemma into other games, compares Prisoner's Dilemmas with other families of games, traces paths that affect the difficulty of transforming Prisoner's Dilemma and other social dilemmas into win-win games, and shows how ties connect simpler and more complex games. Charts illustrate the relationships between the 144 strict ordinal 2x2 games, the 38 symmetric 2x2 ordinal games with and without ties, and the complete set of 1,413 2x2 ordinal games. Payoffs from the symmetric ordinal 2x2 games combine to form asymmetric games, generating coordinates for a simple labeling scheme to uniquely identify and locate all asymmetric ordinal 2x2 games. The expanded topology elegantly maps relationships between 2x2 games with and without ties, enables a systematic understanding of the potential for transformations in social dilemmas and other strategic interactions, offers a tool for institutional analysis and design, and locates a variety of interesting games for further research.*

## Introduction

Cooperation can be difficult. Prisoner's Dilemma, Chicken, Battles of the Sexes, Stag Hunts and the rest of the games where two players each have two strategies (2x2) provide elementary models of conflict and cooperation (Schelling 1960, Rapoport et. al. 1976). One 2x2 ordinal game can be converted into neighboring games by swaps in adjoining payoffs,. The resulting topology, a network of nodes and links, a graph with vertices and edges, elegantly organizes 2x2 games according to their properties (Robinson and Goforth 2005).

The eight possible preference orderings for "non-strict" ordinal 2x2 games (Fraser and Kilgour 1986, Kilgour and Fraser 1988) provide a useful way to organize the complete set 2x2 games with and without ties. These can be linked in an expanded topology of 2x2 ordinal games by half-swaps that form ties (Robinson, Goforth, and Cargill 2007).

Using enhanced visualizations of the topology of 2x2 games, this paper discusses where Prisoner's Dilemma fits within the larger diversity of games, and the pathways through which Prisoner's Dilemmas and other social dilemmas can be converted into win-win games. Just as a prisoner seeking to dig a tunnel to escape his cell might like to have a map, so he can avoid ending up in another bad situation and instead reach somewhere better, the topology of 2x2 games provides a map for understanding how games can be transformed.

The following sections of this paper describe how payoff swaps turn Prisoner's Dilemma into other games, compare Prisoner's Dilemmas with other families of games, trace paths for transforming Prisoner's Dilemma and other social dilemmas into win-win games, and show how ties connect simpler and more complex games in the complete set of 2x2 games.

## Changing Games

Prisoner's Dilemma is not alone. Prisoner's Dilemma is the most famous (or infamous) game, along with its multiplayer equivalents, such as the Tragedy of the Commons (Hardin 1968, 1998) and the Free Rider Problem (Olson 1971). In the story (Tucker [1950] 2001), two prisoners, held separately, are offered a deal where if one confesses and the other does not, the first gets a reward and the other a severe penalty. If both keep silent (cooperate) they will go free, while if both confess (defect) they both receive a milder penalty. If each player follows the logic of their individual incentives, they end up with a worse result than if they could cooperate in staying silent. Cooperation is unstable, since each player has a dominant strategy that means they do better by defecting, whichever strategy the other player chooses. As long as each expects that the other will defect, then the best response Nash Equilibrium leaves both trapped at their second-worst result.

Robinson and Goforth's (2005) topology of 2x2 games provides a framework for understanding the results of changes in payoffs that switch the ranking of different outcomes, and how Prisoner's Dilemma can be transformed into other games. Starting from Prisoner's Dilemma (Figure 1, and the legend in Plate 1), swapping the top two payoffs for the row player creates an Asymmetric Dilemma which shares an equally poor, Pareto-deficient equilibrium. Swapping the middle two payoffs for the same player produces a game with a Pareto-deficient outcome where one player gets second best and the other third best, even though there is a Pareto-optimal outcome where the first player could get their best outcome and the other second best. Such asymmetric situations could occur if one player has an alibi, and so faces a lesser penalty (Robinson and Goforth 2004, 2005). A further swap in the lowest two payoffs creates another game, Revelation (Brams 1994:103-11) with a similarly poor, Pareto-deficient equilibrium.[2]

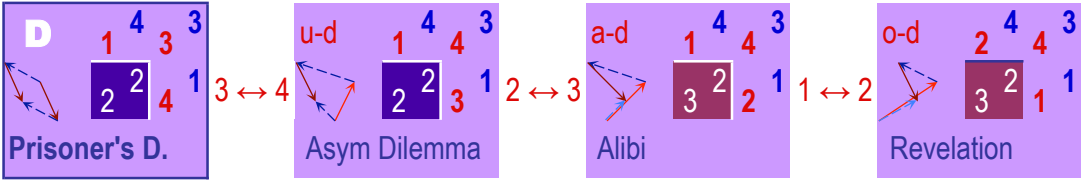

*Figure 1: Prisoner's Dilemma turns into Asymmetric Dilemma, Alibi, and Revelation, all with Pareto-deficient outcomes.*

Prisoner's Dilemma has other neighbors. Swapping the middle two payoffs for either player creates a game of Total Conflict, a fixed rank-sum game (the ordinal equivalent of zero-sum) where any improvement for one player is matched by an equivalent worsening for the other, with an equilibrium where one player gets second-best and the other second-worst (Figure 2). A second swap of lowest ranked payoffs for the same player yields a game with similarly tragic results, and still without even the potential for a Pareto-superior result.

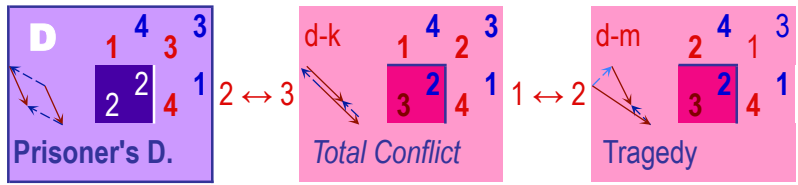

*Figure 2: Prisoner's Dilemma turns into Total Conflict, and then Tragedy.*

From Prisoner's Dilemma, swapping the lowest payoffs for one person creates a game of Called Bluff with an equilibrium where the other player has a dominant strategy that leads to their best outcome, and the first is stuck with their second-worst result (Figure 3). Swapping the lowest payoffs for the other player then creates a game of Chicken. In Chicken, the second most famous game (also known as Hawk-Dove), if both choose Dove strategies, each could get their second best result, but each would then have an incentive to defect and get their top preference. If both play Hawk, both get their worst outcome. The combinations where one plays Hawk and one plays Dove create two Nash Equilibria, where given the other player's strategy, neither player can act on their own to unilaterally improve the situation; although one player gets their best result and the other their second-worst.

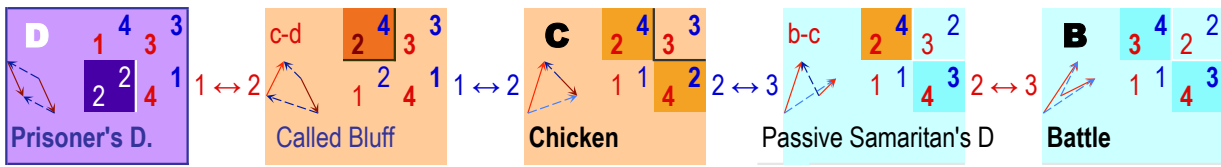

*Figure 3: Prisoner's Dilemma turns into Called Bluff, and into Chicken, and further transforms into an Asymmetric Battle of the Sexes and then Leader, a proper Battle of the Sexes.*

A further swap in the two middle payoffs transforms Chicken into an asymmetric Battle of the sexes, which Buchanan (1975) analyzed as Passive Samaritan's Dilemma. A further swap in middle payoffs for the other player results in the Battle of the Sexes game that (Rapoport 1967) called Leader. Like Chicken, there are two equilibria, in this case combining best and second-best payoffs for the players, so each prefers a different equilibrium, and they face the risk of both doing worse if they cannot coordinate.

As discussed by Robinson and Goforth (2005), changes in payoffs, and resulting swaps in rankings, could come from "*small* changes in information, preferences, or technology, or from small errors in identifying games" [emphasis in original]. Swaps in payoff ranks could result from a variety of sources including random disturbances (trembling payoffs), better information, greater consideration of impacts on the other person (other-regarding preferences), changes in technology and organization that reshape production functions, revised rules, changes in the broader institutional context, shifts in monitoring that affect expected benefits and costs (including the likelihood of penalties or effects on reputation), or deals creating side payments between players. In contrast to treating preferences as fixed, the topology of swaps in payoffs provides a systematic way to look at the impact of changes in preference rankings. To the extent that shifts in payoffs can be deliberately shaped by those involved, the topology of 2x2 games may not only help to understand institutions and their consequences, but also be a useful tool for institutional design.

## An Elegant Diversity of Games

Any strict (no ties) 2x2 ordinal game can be converted into six other ordinal games by switching adjoining payoff ranks: the lowest two (1↔2), middle two (2↔3), or highest two (3↔4), for either player. The topology of 2x2 games (Robinson and Goforth 2005) provides an elegant and practical way of organizing and understanding the set of games that result from swaps. Starting from Prisoner's Dilemma, or any other game, the topology can be generated by first swapping the two lowest values to form a *tile* of four games that are closest to each other in the topology. Thus, Prisoner's Dilemma is in the same tile as Chicken and two versions of Called Bluff (Figure 4). Swaps in the middle payoffs, followed by swaps in the lowest payoffs and further middle swaps, generate new tiles, resulting in a *layer* of nine tiles and thirty-six games. The layer is a torus, since further swaps return to the same games.

Figure 4. Prisoner's Dilemma Tile

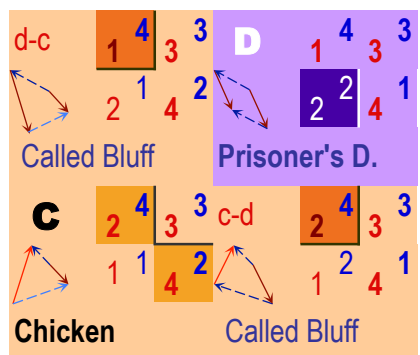

From Prisoner's Dilemma, (or any other game,) swapping the highest two payoffs begins another layer, which can be completed by further low and middle swaps. There are four layers, which differ by the alignment of the top two payoffs. Harmony and other win-win or "no-conflict" games with the highest payoffs in the same cell form one layer, designated by Robinson and Goforth as Layer Three. Highest payoffs are diagonally opposite each other on what Robinson and Goforth designated as Layer One, which in this paper may also be referred to as the Discord Layer, and which includes Prisoner's Dilemma, Chicken, and the Battles of the Sexes. The other two layers, to the left and right, have the top two payoffs aligned in the same column or row respectively. Each layer is a torus, therefore making it possible to scroll Prisoner's Dilemma to the center, which conveniently organizes games according to dominant strategies and other properties. This splits up the Prisoner's Dilemma tile so it lies on the corners of each layer, and shows many of the most interesting high swaps between layers, since the entire table is then also a torus connected on the edges. Plate 1 shows an enhanced visualization of the topology, based on Robinson and Goforth's (2004) Periodic Table of 2x2 Ordinal Games, with additions including numeric payoffs, payoff families, and a simple naming scheme for uniquely identifying 2x2 ordinal games.

The symmetric games, where each player faces the same payoffs, form an axis from southwest to northeast (Plate 1 and Plate 2a). Each row has the same payoffs for the row player, and each column has the same payoffs for the column player. The asymmetric games on either side of the axis are mirror reflections of each other, swapping positions for row and column. If reflections are treated as the same game, then the twelve symmetric games, and sixty-six games on one side of the axis add up to the total of 78 unique games identified by Rapoport and Guyer (1966, 1976). However, understanding the topological structure of how games are related by swaps in adjoining payoffs requires using the full set of 144 strict ordinal games.

The 144 games can generated by combining the twelve possible payoff patterns for each player. A simple, Cartesian-style convention puts Row's highest payoff in the right-hand column and Column's highest-payoff in the upper row[3], so each game also represents variants formed by switching rows and/or columns. Any asymmetric game can be located as the combination of payoffs from two symmetric games.[4] The resulting topology shows the pathways through which games can be transformed. It puts Prisoner's Dilemma into the context of the larger number of games that are mostly asymmetric, and mostly have much better results.

The actual distribution of incentive structures in social situations is an interesting question for empirical study. From a theoretical perspective, the 144 games in the topology display all possible combinations of four ranked payoffs for each of the two players.[5] If payoffs are generated randomly (and not restricted to a few discrete values) and then converted into games with four ordinal ranks, the resulting games will tend toward the proportions in the strict ordinal games (Simpson 2010:56, 2011:10). Thus, to the extent that payoffs arise randomly, the proportions of the 144 strict games in topology of 2x2 games provide a basis for default expectations of how likely different game situations are to occur.

Table 1. Distribution of Nash Equilibrium Payoffs in 2x2 Games

| Payoff Families and Subfamilies | Sym-metric | Asym-metric | Total | % | Nash Equi-libria | Dominant Strategies | Pareto Optima |
|---|---|---|---|---|---|---|---|
| **1. WIN-WIN 4-4** | **6** | **30** | **36** | **25%** | **1** | **1-2** | **1** |
| Harmonious | 3 | 24 | 27 | 19% | 1 | 1-2 | 1 |
| Stag Hunt | 3 | 6 | 9 | 6% | 2 | 0 | 1 |
| **2. BIASED 4-3** | **2** | **42** | **44** | **31%** | **1-2** | **0-2** | **2** |
| Altruistic | 0 | 12 | 12 | 8% | 1 | 1 | 2 |
| Altruistic & Self-serving | 0 | 12 | 12 | 8% | 1 | 2 | 2 |
| Self-serving | 0 | 12 | 12 | 8% | 1 | 1 | 2 |
| Battles of the Sexes | 2 | 6 | 8 | 6% | 2 | 0 | 2 |
| **3. SECOND BEST 3-3** | **2** | **10** | **12** | **8%** | **1** | **1-2** | **3** |
| **4. UNFAIR 4-2** | **1** | **18** | **19** | **13%** | **1** | **1-2** | **2-3** |
| Chicken | 1 | 0 | 1 | 1% | 2 | 0 | 3 |
| Winner | 0 | 6 | 6 | 4% | 1 | 1 | 2-3 |
| Win-Lose | 0 | 6 | 6 | 4% | 1 | 2 | 2-3 |
| Loser | 0 | 6 | 6 | 4% | 1 | 1 | 2-3 |
| **5. PD FAMILY - Extended** | **1** | **14** | **15** | **10%** | **1** | **1-2** | **2-3** |
| Prisoners' Dilemma 2-2 | 1 | 2 | 3 | 2% | 1 | 1-2 | 2-3 |
| Alibi 2-3 | 0 | 4 | 4 | 3% | 1 | 1 | 2 |
| Tragic 2-3 | 0 | 8 | 8 | 6% | 1 | 1-2 | 2-4 |
| **6. CYCLIC** | **0** | **18** | **18** | **13%** | **0** | **0** | **2-4** |
| | 12 | 132 | 144 | 100% | | | |

While Prisoner's Dilemma has received by far the most attention from researchers, it and the six other games with Pareto-deficient equilibria actually compose only a small proportion of games, about five percent. Even if reflections are counted as equivalent, Prisoner's Dilemmas still only make up 4/78=5.13%. So, the other 95% of games are not Prisoner's Dilemmas. There could be situations and selection processes that tend to create Prisoner's Dilemmas. However, in practice it seems likely that there will also be selection against Prisoner's Dilemmas, as people avoid playing such games or seek to change them into games with better results.

Win-win games, make up one fourth of the games. So, cooperation in social life may be easier to achieve and more common than suggested by the disproportionate attention to Prisoner's Dilemma. And, although game theorists have sometimes labelled games in which both players get their best result as (mathematically) "trivial" or "boring," for those involved, successful cooperation is often vital.

The subset of win-win games with two equilibria, the Stag Hunt games, pose special challenges to achieving cooperation (Skyrms 2004). In a review article, Kollock (1998) conjectured that Assurance games might be more common than Prisoner's Dilemma. All nine of the strict ordinal Stag Hunt games, with two Nash Equilibria, one Pareto-inferior, fulfill Sen's (1967) definition of an assurance game: both players would do best by cooperating, but if one does not cooperate then the other is also better off also choosing the non-cooperative strategy. If payoffs occur randomly, the nine assurance games would be slightly more frequent than the seven Pareto-deficient games, confirming Kollock's conjecture.

The zero-sum games, where any gain for one player is matched by an equivalent loss for the other, were the focus of much of early game theory, including Von Neuman and Morgenstern's foundational work (1947). The ordinal equivalents of zero-sum games, with a fixed rank-sum for both players, form a diagonal line oriented northwest to southeast near the center of the topology display, including Zero Sum, Big Bully and Total Conflict and their reflections). The six fixed rank-sum games make up less than five percent of the total. Thus, nonzero-sum games, those without a fixed rank sum, make up over 95% percent of the strict 2x2 ordinal games.

The topology also provides a way of looking at the distribution of games of pure conflict, pure cooperation, and mixed interests (Schelling 1960, Greenberg 1990, Robinson and Goforth 2005). The fixed rank-sum games are part of the larger set of games of pure conflict, where each player's incentives always induce them to make moves that reduce the payoff for the other player, creating a negative externality (Plate 2d). These sixteen games make up just over 11% of games. The fourteen games of pure cooperation, where one player's incentives always lead them to make things better for the other, are slightly less frequent, just under 10%. Robinson and Goforth (2005) also identified eight "Type" Games, where the inducement correspondences (Greenberg 1990) are always positive for one person and always negative for the other; one person's incentives always lead to positive externalities and the other person's incentives always lead to negative externalities. In more colloquial terms, incentives make one person kind and the other cruel, suggesting the name Jekyll-Hyde Type games. And, in unfortunate contrast to the usual outcome in stories and movies, in the Jekyll-Hyde Type games the good guys lose and the villains triumph.

Somewhat surprisingly, the largest family of games, categorized according to payoffs at Nash Equilibria, are those with biased payoffs combining the first and second best outcomes, as in the Battles of the Sexes. Given the variety of possible joint incentive structures, it is not accurate to assume that Prisoner's Dilemma is the only or most likely model of a social situation with strategic interaction between two actors. In terms of the variety of possible games, and their distribution, if payoffs are generated randomly, almost two-thirds of games (92/144=64%) have Nash Equilibria that allow both players to get their best or second-best payoff.

The topology of 2X2 games provides a way to escape from the assumption that Prisoner's Dilemma is the most prevalent or typical form of strategic interaction, and instead understand the diversity of different incentive structures that may actually be present. To the extent that payoff patterns occur randomly, two other kinds of social dilemmas,

Assurance games and Battles of the Sexes, where uncoordinated action also risks resulting in Pareto-inferior outcomes, would be more common than Prisoner's Dilemmas. Although most research has concentrated on symmetric games, asymmetric games are far more numerous.

Furthermore, the majority of games have unequal outcomes, as in the Biased, Unfair, Tragic, and Alibi Games. To the extent that payoffs occur randomly, equality in outcomes cannot be assumed or expected to occur automatically. The likelihood of asymmetric structures and unequal outcomes justifies studying the dynamics of inequality, as well as the opportunities for making changes that might lead to more equitable results.

## Paths to Win-win

In the classic story for Prisoner's Dilemma, the offer to reduce the penalty for the prisoner who confesses can be seen as a form of social engineering, deliberately rearranging payoffs so there is an incentive to confess, with the added twist of a greater reduction if only one confesses. The dilemma comes from the structure of payoffs for both of the strategies for the other person: if the other stays silent, it is better to confess, and if the other confesses it is still less bad to also confess.

If the other person might retaliate against defection, motivated by revenge, maintaining a reputation, or other reasons, this could change the expected payoffs, possibly making keeping silent a better option if the other keeps silent. This would switch the two highest payoffs, a 3↔4 swap. If the person threatening retaliation would still themselves be susceptible to the temptation to defect, the situation becomes an Asymmetric Dilemma, in which the Nash Equilibrium is still the Pareto-deficient outcome, since one would defect and then the other is better off by defecting (Figure 5). If the threat of punishment for defection was mutual, then the game would become a Stag Hunt, where if one trusts the other to cooperate, then both can achieve their best result. However, if each fears that the other will not cooperate and instead plays to minimize risk, avoiding the worst outcome, then even in a Stag Hunt, risk-avoiding strategies end up at the second-worst payoffs.

A norm of silence, such as the Sicilian *omerta*, perhaps backed by rewards and punishments from an organized gang, would also amount to a way of adjusting expected payoffs. The threat of punishment, in or after prison, would make defecting less attractive, compared to cooperation, and so could switch the ranking of the top two outcomes. Even if the other defects, the promise of support, for a prisoner and his family during and after a prison sentence, could switch the ranking of the lowest two payoffs. The combination of swaps in the highest and lowest payoffs would transform Prisoner's Dilemma into the highly stable game of No Conflict, where both have dominant strategies to cooperate in keeping silent. Without the added twist of a greater reduction in sentence if only one prisoner confesses, the situation in Prisoner's Dilemma could be a Stag Hunt, which the prosecutor is trying to modify with a tricky offer of less punishment if one prisoner defects and the other does not.

Rather than being trapped in a single rigid payoff structure, there are options available for shifting payoffs, and the accompanying structure of incentives.  As another example, a witness protection program offers a way to safeguard the incentives to defect, albeit one that is expensive and perhaps less than completely certain. Transformations between games provide simple "toy" models of how changes in payoffs restructure incentives and strategic interactions. The story of Prisoner's Dilemma is a simple way of describing the consequences of such an incentive structure and its potential transformations, which can apply to a much broader variety of situations.

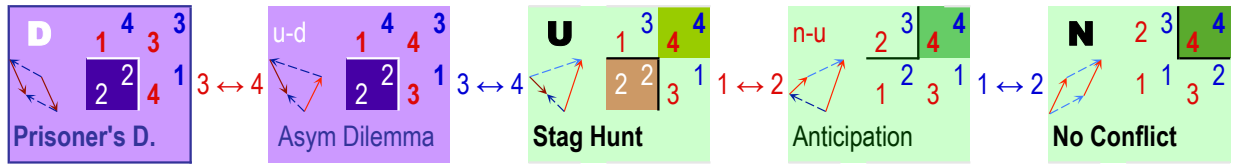

*Figure 5. Prisoner's Dilemma turns into an Asymmetric Prisoner's Dilemma and then transforms to a precarious Stag Hunt and then a Privileged Hunt, which could then become a game of No Conflict.*

The common "solution" to Prisoner's Dilemma is through repeated play, in which case the Folk Theorem shows that cooperation is a Nash Equilibrium in the iterated supergame, at least if there is no definite end to the game. In this case, the expected payoffs from cooperation are higher, since it also makes sense for the other to cooperate. This can be supported by strategies such as Tit-for-Tat, cooperating on the first stage and then doing whatever the other person did last time (Axelrod 1984). However, if the other person cannot be trusted to cooperate, then a risk-minimizing strategy could still avoid the worst payoff, creating a second Nash Equilibrium, in which case the expected payoffs would be those of a Stag Hunt, an assurance problem where choosing the  cooperative  strategy when the other does not is worse than defecting.

The solution for repeated play does not solve the game if it is only played once (Binmore 2007). The best option for dealing with single-play Prisoner's Dilemma is to avoid the game or change it, either by not getting into such a game in the first place, or by finding ways to change the payoffs. Social norms and rules, backed by informal or formal sanctions, represent ways of changing payoffs, making cooperation more beneficial and defection more costly, and so transforming incentive structures to yield better results. The transformations discussed here illustrate, in simple, elementary models, some of the options for changing the game.

In dealing with free-rider problems, a multiperson equivalent of Prisoner's Dilemma, where each individually finds it better to not contribute, Olson (1971) argued that an important form of solution was through "selective incentives" that tied a non-excludable public good to some excludable good. Thus a professional association might offer a journal, insurance, or other benefits only to members. This raises the payoff associated with cooperating compared to defecting, as long as the others cooperate. The two-person equivalent to switching the relative ranks of these two payoffs coverts the Prisoner's Dilemma into a Stag Hunt.

For "strategic moves" within a game, a player who could credibly commit to a particular strategy could force the other player to go along. The most famous example, is "throwing away the steering wheel" when two cars play Chicken, thereby forcing the other driver to swerve (Schelling 1960). Narrowing choices down to a democratic vote between two alternatives is another way to avoid the conflicts that may stymie agreement in the face of more complex incentive structures (Brams and Kilgour 2009).

In contrast to strategic moves *within* games that *reduce* options, the potential for changing the game, through rewards and punishments, norms, rules, side payments, and other changes that affect expected payoffs, provides a way of understanding possible solutions to social dilemmas, making strategic moves *between* games.[6] The feasibility of such changes in payoffs would depend on the specifics of a particular situation, in the same way that "throwing away the steering wheel" would depend on having a steering wheel that is easily detached and which can be conspicuously disposed of so the other person actually believes that there is no alternative, or, more realistically, some equivalent commitment mechanism. For example, a politician or faction might make a public statement locking themselves into an extreme, non-cooperative strategy, in an attempt to force others to concede. Solutions resulting from repeated play, and the expected payoffs from future games and the benefits of a good reputation also depend on the details of a particular social context, and the knowledge that people bring to it. However, the key point is that in many cases, changes can be made that would change the expected payoffs from a game and the ranking of outcomes, transforming the incentive structure and equilibrium results.

The topology assumes games linked by swaps in the lowest two payoffs are closest to each other. Similarly, it could be assumed that, other things being equal, swaps in lower-ranked payoffs would have lower transaction costs than swaps in higher-ranked payoffs. An assumption about which swaps are likely to have lower transaction costs provides a way of identifying paths that might be more easily available and more likely to be pursued.

Within the topology, Chicken is close to Prisoner's Dilemma, part of the same tile of games that differ only in the arrangement of the lowest two payoffs. However, the potential transformations available from the two games differ. Swaps in the highest payoffs for one person change Chicken into a game of Hegemony. In this case, one player has a dominant strategy, but ends up doing much worse than the other. The reluctance to be trapped in such a situation may contribute to the stability of Chicken. However, a further change in the ranking of the top two payoffs for the other player would then create a game of No Conflict (Figure 6). So, from Chicken, there is a much shorter and more direct pathway to a harmonious win-win game than is available from Prisoner's Dilemma, if the rankings for both players can be changed.

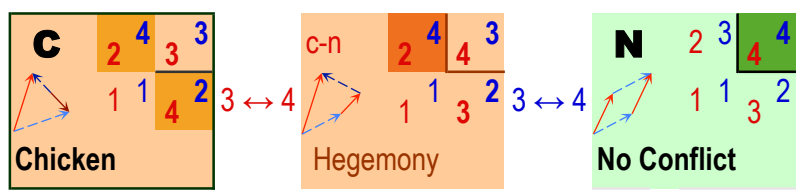

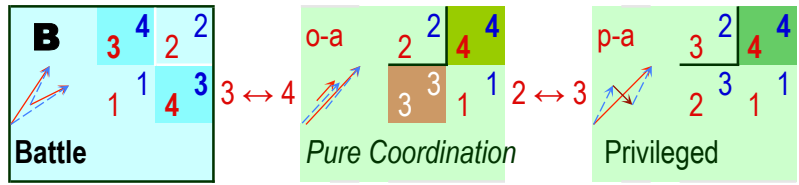

*Figure 6. Chicken turns into Hegemony, and then to No Conflict.*

Layers are linked by swaps in the two highest values, such as the transformation from Chicken into Hegemony and No Conflict. The four proper Battle of the Sexes games on Layer One turn into the four proper Coordination Games on Layer Three. Any game on the Battle of the Sexes tile is linked by row or tile swaps to two games on the Coordination tile, and vice versa, forming what Robinson and Goforth call a *hotspot (Plate 2d)*.[7] For the games of Coordination, Pure Coordination and Asymmetric Coordination a further swap in middle payoffs creates a Privileged, dominance-solvable, game, where one player has a dominant strategy and, based on that, the other player can figure out her best choice of strategy, and both achieve their best result (Figure 7).

*Figure 7: Battle of the Sexes (Leader) turns into Pure Coordination, and can then turn into a Privileged game. (The columns are interchanged to keep Row's 4 right)*

In the case of the classic Battle of the Sexes story of wanting to do something together, but having conflicting preferences, communication might enable one person to persuade the other to change preferences, whether coming to appreciate the drama involved in a boxing match or the struggles and adventures of an opera. An increase in "other-regarding preferences" could make being together more important than attending one's favorite kind of event. The classic story (from before the days of cellphones) has the couple unable to communicate and trying to figure out where to meet for the evening. The conflict in preferences can occur even where communication is possible, but communication opens many more options for trying to shift preferences, through persuasion or various kinds of rewards, threats, and other deals. While these could be included in more elaborate supergame models, the shift between games captures the essential change in preference structures that creates a new, stable outcome.

In Buchanan's (1976) Samaritan's Dilemma, one person can make things better for themselves even without aid, likes it better if they also get help, but likes it best if they can get aid while making less effort themselves. Since the donor always prefers the outcomes with aid, they are trapped in the Samaritan's Dilemma, getting only their second-best outcome. Unlike Prisoner's Dilemma, repeated play does not provide a solution to Samaritan's Dilemma (Schmidtchen 1999). Buchanan's proposed solutions depend on deceptive threats or enforcement by a third party, If there is no binding third-party enforcement, then this is basically a bluff by the donor to threaten that they will not provide aid unless the recipient also invests.

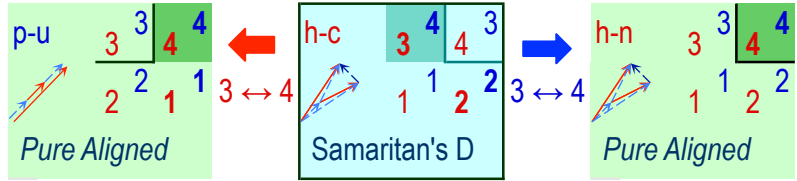

*Figure 8: Samaritan's Dilemma can turn into either of two win-win, Pure Aligned games.*

Shifting payoffs so that effort by the recipient, when there is aid, is more rewarding for the recipient, converts Samaritan's Dilemma into a win-win game. Bureaucratic aid may be delivered in ways that discourage joint effort, so changes that make joint investment easier represent a way of transforming the incentive structure. For example, in government irrigation or other forms of investment, this might be a matter of simplifying requirements, or effectively using matching formulas (Bruns 2010).

A closer examination of many apparent "Samaritan's Dilemmas" may help to understand the efforts already being made by potential recipients, the benefits of not requiring excessive sacrifice by them, and the benefits donors themselves receive from giving assistance (Stone 2008). Donors might come to accept a greater need for some kinds of social support, rather than strict, and perhaps unrealistic, expectations regarding self-reliance. Such changes might switch the ranking the donor assigns to the two outcomes in the case of assistance, converting Samaritan's Dilemma into another win-win game. The two possible transformations (Figure 8), by row or column swaps, from Samaritan's Dilemma are part of the hotspot that links Samaritan's Dilemma and the other games on the same tile with the equivalently located games on the Win-win Layer, transforming Samaritan's Dilemma into one of two forms of a Pure Aligned game, where both players have the same values for each cell.

While Robinson and Goforth explain the topology of 2x2 games in terms of swaps in adjoining payoffs, the topology can also be applied as a framework for looking at swaps in non-adjoining payoffs, which might result from a drastic increase or decrease in one or more payoffs. Buchanan (1977) used a 2↔4 swap to switch between active Samaritan's Dilemma, *h-c*, with its single Nash Equilibrium, to Passive Samaritan's Dilemma, *b-c* (next to Chicken), with two equilibria, but still no sure escape from the Samaritan's Dilemma. Switching the lowest and second-highest payoffs (1↔3) stays on the same layer, and moves to one of the adjoining tiles. Thus, Prisoner's Dilemma would turn into a Patron game, *r-d*, in which the player with a dominant strategy does best, while the other gets second-best. Swapping a four switches layers, so a 2↔4 swap from Prisoner's Dilemma puts the top payoffs in the same column, moving to a Generous game, *h-r*, on Layer Two. Swapping the lowest and highest payoffs puts the highest payoffs into the same cell, moving to a Pure Aligned game, *h-n*, on the Win-win layer.

Swaps in adjoining payoffs create six neighbors for each game, and double or triple swaps create six more distant neighbors. One could assume that, other things being equal, swaps in non-adjacent payoffs would be less likely, or involve higher transaction costs. Mathematically, a double or triple swap can be seen as composed of repeated swaps in adjoining payoffs, just as multiplication is repeated addition. However, empirically, various changes in technology, profitability, likelihood of detection and punishment, or other factors could produce a big jump, catapulting one outcome beyond others.

While it may be easy to look at one or two swaps for a particular game, the topology provides a systematic way to understand the full set of potential transformations. In particular, the topology of 2x2 games maps pathways to win-win. At the same time it also reveals the paths away from win-win, vulnerabilities, illustrating the ways in which win-win games risk ending up with unequally ranked outcomes. Thus Coordination games could turn into Battles of the Sexes, a harmonious Pure Aligned game could turn into a Samaritan's Dilemma, No Conflict could turn into Hegemony, and Stag Hunt could turn into Prisoner's Dilemma. An important aspect of such vulnerability is what Williamson (1996) calls post-contractual opportunism. Unanticipated changes and the inherent incompleteness of contracts and their enforcement may mean that even when incentives were initially well aligned, things come undone, creating a situation where one or both parties may be disadvantaged or tempted to defect from cooperation.

## Ties Make Simpler Games

The basic situation in Prisoner's Dilemma, before the offer of a reduced penalty in exchange for a confession, might be summarized in a very simple game, where both do best if they cooperate in remaining silent, and otherwise receive the same penalty whether one or both confess. In such a case, the weakly dominant strategy for both is to keep silent. The prosecutor's offer creates a more complex game, with different payoffs for each of the four combinations of strategies. This is equivalent to moving from a very simple game, here called Basic Harmony, with one best outcome for both players who are indifferent between the other three outcomes, to the more complex situation of Prisoner's Dilemma.

From Prisoner's Dilemma, ties in the *lowest* two payoffs create a *borderline* game intermediate between Prisoner's Dilemma and Chicken (Rapoport et al. 1976:31), which here will be called Low Dilemma. This game can be seen as lying at the center of an expanded Prisoner's Dilemma tile (Robinson, Goforth and Cargill 2007), as shown in Figure 9. Each player still has an incentive to defect from the cooperative outcome, and when both act accordingly, they both receive the worst payoff. Neither can unilaterally improve their payoff, and so in this sense it is a best response equilibrium. On the other hand, neither has anything to lose by deviating from the poor "equilibrium," and could gain if the other makes a mistake or for any other reason acts cooperatively. So, equalizing the two lowest payoffs, or changes that make them indistinguishable, for example due to uncertainty about whether promises will be fulfilled, does not completely escape from Prisoner's Dilemma, but may make the Pareto-inferior outcome less stable.

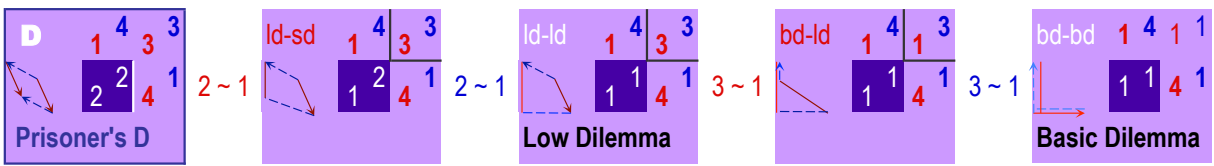

*Figure 9. Expanded Prisoner's Dilemma Tile with Low Dilemma in Center*

As long as being the only person to confess yields the highest payoff, the conflict remains. Setting the three lowest values to be the same would yield the most basic game of conflict, Basic Dilemma (with ties for the lowest three payoffs, 1~2~3), as in Figure 10. The best results are diagonally opposite each other, on opposing strategies for each player, capturing the essence of the conflict. This game is the archetype or exemplar for Prisoner's Dilemma, Battles, of the Sexes, and the rest of the games on the Layer One. For single play, there is no way both can win, and if each chooses the strategy that includes their highest payoffs, then both get nothing. However, if play is repeated, then taking turns would be an obvious solution, if they can somehow synchronize their choices.

*Figure 10. Making ties (half-swaps) turns Prisoner's Dilemma into Low Dilemma and then Basic Dilemma*

Looking at the four games in the center of the chart (Plate 1 and Figure 11), a half swap in the *highest* values for one player creates a game, halfway between Prisoner's Dilemma and Asymmetric Dilemma. This would happen if the benefits of being the only person to defect are reduced, or the benefits from successful cooperation are increased. This might occur due to uncertainty about promises for leniency being fulfilled or fears of retribution for confessing. This could create a situation where the player cares about the differences between the two lowest payoffs, and so prefers to confess if the other confesses, but, if the other cooperates in keeping silent, is indifferent between the cooperative or defection outcome, having lost the incentive to defect from cooperation. A half swap for the other player makes this situation symmetric, creating a High Dilemma, (creating ties between 3s and 4s, 3~4). The win-win outcome is a Nash Equilibrium, best responses from which neither has an incentive to defect. However, the weakly dominant strategy is still to defect, and a safety-first, precautionary, maximin strategy of avoiding the worst outcome would also lead to the same poor result.

As a variation on Rousseau's story, in the High Dilemma if the other person acts cooperatively, then the first person gets their best payoffs regardless of which strategy she chooses, for example, if the other person helps to "beat the bush" and drives enough game toward the first person that she expects to do as well as with her share of the stag, e.g. from lots of rabbits or other small game. As a variation on the Prisoner's Dilemma story, as long as the other person stays silent, does not defect, then the first player still expects to get the same payoff whether she confesses or not (perhaps not believing the promised reward). However, if the other does confess, then it is better to also confess. In either case, this creates a situation poised between a Prisoner's Dilemma and a Stag Hunt. The High Dilemma combines a weak form of Prisoner's Dilemma with the assurance problem of a Stag Hunt, making the win-win outcome even more difficult to achieve than in the Strict Stag Hunt. As a borderline game, High Dilemma sits at the boundary between the win-win and discord layers, and between the Stag Hunt games and the Prisoner's Dilemmas.

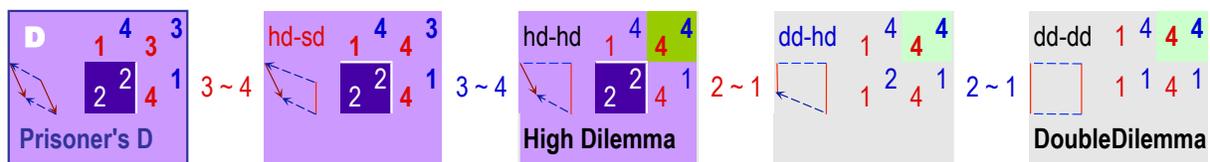

*Figure 11. High half swaps turn Prisoner's Dilemma into High Dilemma and then low half swaps create the Avatamsaka Game*

From the High Dilemma, half swaps equalizing the lowest two payoffs for both form an even simpler game, where each player has two ties, the Avatamsaka Game, (Aruka 2001, cited in Goforth and Robinson 2011), here labelled double dilemma, which can be seen as an example of interdependence. For each of the other person's strategies, a player gets the same payoffs, what Rapoport et (1976) call a "degenerate" game, and there is no dominant strategy solution. However, if each player does what will be best for the other, then both can reach the win-win outcome. If each player expects the other to play cooperatively, then neither can improve their payoff by defecting, so creating a best response equilibrium. The game exemplifies the Golden Rule, the ethical principle found in many religions, to treat others as you would like to be treated. In more game theoretic terms, the win-win outcome is a prominent focal point, and neither player has an alternative that yields a higher or safer payoff, and so, arguably, it makes sense to choose the strategy that might lead to the best result.

Starting from Chicken, a half swap in middle payoffs creates a game between Chicken and an Asymmetric Battle of the Sexes. A half swap for both players creates Volunteer's Dilemma (Poundstone 1992) or Middle Battle. This game can model a situation in which one person's action is sufficient to provide the shared benefit, and adding a second contribution makes no meaningful difference in the value of the result (for example a second call to 911), and each prefers that the other person does the work. The risk is that neither acts, leading both to their worst outcome. One path to a solution to Volunteer's Dilemma would be reflected by swapping the top two payoffs between the equilibrium outcomes, so that one person's top preference is to be the sole volunteer, is strongly averse to not calling if the other does not call, and is indifferent between helping or not helping if the other person is helping.[8]

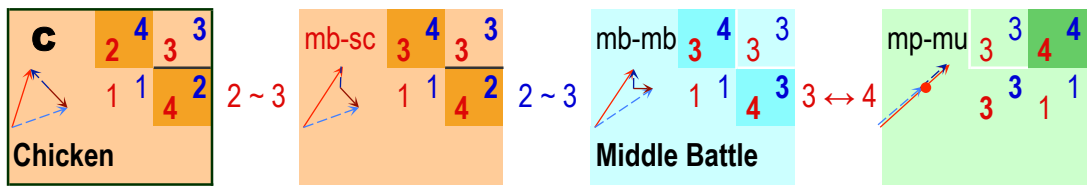

*Figure 12. Chicken turns into Volunteer's Dilemma (Middle Battle)*

The examples above illustrate some specific transformations through half swaps that make ties. For a more general understanding of transformations that make simpler games, it would be useful to have a systematic way to display and identify the full set of non-strict games. For each person, there are eight possible preference orderings, from utter indifference to four strictly ranked preferences (Fraser and Kilgour 1986, Kilgour and Fraser 1988). This yields a total of 64 preference classes, (Plate 3a). As discussed by Robinson, Goforth, and Cargill (2007), ties can be seen as half swaps that form a game "between" two existing games. Thus the eight preference orderings are created by the half swap operations that can be carried out on a strict game to form ties, first forming high, middle, or low ties, and from there high and low (double), three highest (triple), or three lowest ties (basic), and finally all ties (zero). The preference classes and topological links of swaps and half-swaps then provide a "natural ordering" that can be used to arranged the complete set of 2x2 games.

The strict games make up the largest class, 144 games out of a total of 1,413 distinct non-strict games, if reflections that swap positions are treated as distinct games (if reflections are treated as equivalent, then there are 726 unique ordinal games, Guyer and Hamburger, cited in Robinson et al. 2007). The games with ties can be located within the remaining 63 classes, first categorized according to the number of different payoff ranks (four, three, two, or one). With three payoff ranks, there may be ties for the highest, middle, or lowest ranked payoffs. With only two payoffs, there may be ties for the three lowest or three highest payoffs, or ties for the two lowest and two highest payoffs. Robinson, Goforth, and Cargill explain how a topology of "nonstrict" 2x2 games can be formed, although they do not provide a display of the resulting complete topology, which is provided here.

As discussed earlier, any strict asymmetric game can be located as the combination of payoffs from two symmetric games. Thus, the strict cyclic game of Fixed Sum combines payoffs from Assurance and Hero, *A-R*. In the same way, any non-strict asymmetric game can be located as the combination of payoffs from two symmetric ordinal games. For example, Jekyll-Hyde, *mh-mk,* the simpler version of the strict Jekyll-Hyde Type games (Robinson and Goforth 2005) is located in the class of games with ties in middle payoffs and is formed by combining the payoffs from the symmetric games of Middle Harmony, *mh,* and Middle Deadlock (Midlock),[9] *mk*. This is the simplest game in which one player always has incentives to increase the other's payoffs, while second player's incentives lead to reducing the payoffs to the first player, one is induced to be kind, and the other to be cruel.

The full set of 2x2 ordinal games can be generated by combining the payoffs from symmetric ordinal games, including those with ties (Plates 3 and 4).[10] This forms a landscape of social situations traversed by swap and half-swap changes in payoffs. Full swaps move around within a class, such as the class of Strict games, or the Low Ties games. Half-swaps that form ties move between classes, keeping one row or column unchanged while shifting to another class for the person whose payoffs change to make (or break) a tie. Plate Three schematically shows the ordinal games according to the Fraser-Kilgour preference classes. This identifies the thirty-[11]eight distinct symmetric ordinal games, which form a diagonal from lower left to upper right in the display (Figure 3d).

Figure 4 shows what Rapoport, Guyer, and Gordon (1976) call the "complete set" of all possible ordinal 2x2 games, split into four panels. As explained by Robinson, Goforth, and Cargill (2007), the games with low ties and middle ties can be visualized as lying between the strict games in a checkerboard pattern, and this concept is applied to arrange the upper right quadrant of the display. The lower-left quadrant shows the simple two-rank games, which Robinson, Goforth, and Cargill refer to as "archetypal," along with the high ties games formed from pipes and hotspots. The upper-left and lower-right quadrants then show the remaining combinations. Due to the asymmetry between pipes and hotspots, the high ties and double ties games form irregular matrices, from which duplicate games, equivalent by interchanging rows and columns, have been removed.

Having a visualization of the complete set of 2x2 ordinal games facilitates locating games that represent important existing concepts, as well as other potentially interesting games. While most game theory models have concentrate on symmetric strict ordinal games, the elementary dynamics of other situations may be better represented by asymmetric games and their transformations. For example, a simple model of when a principal would offer an agent a contract, expecting the agent to comply, transforms into a moral hazard situation if the agent prefers to defect, swapping the top two payoffs (Figure 13). The principal's best result is a successful contract, and worst if the agent defects. The agent always prefers a contract, is indifferent between the two alternatives without a contract, and the two games differ depending on whether the agent prefers to comply or defect from the contract. The Appendix further discusses games with ties, including names, properties, and interesting games.

*Figure 13. A high swap transforms Principal-Agent cooperation, an asymmetric win-win game with middle and low ties, into a Moral Hazard situation with a Pareto-deficient Nash Equilibrium*

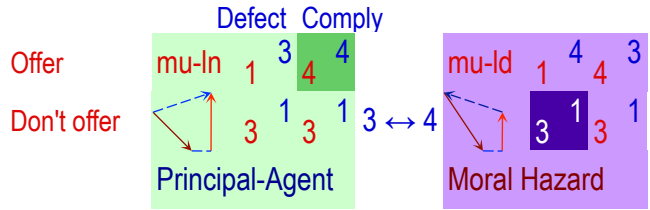

It is also possible to identify larger regions of connected and similar games in the expanded topology. A pathway through the topology (Figure 14) links the strict cyclic game of Fixed Cycle, *A-R*, more completely labelled as*, sa-sr*, to Matching Pennies, the simplest cyclic game.[12] A tie between Column's two lowest payoffs in Fixed Sum would first form a game, halfway between it and the adjoining game of Tragic Cycle, in the class that combines strict payoffs for Row and Low Ties for Column, *sa-lb*. A half swap for Column would move to Low Ties for both, forming a cyclic game, Right Cycle, *lo-lb* , in the center of the tile. Each of the games in the Low Ties class lies at the center of a tile in the strict topology, and many of them form idealized versions of the games in the tile. A further half-swap of Column's highest payoffs would change to the class combining double ties for Row and low ties for Column, *do-lb*. A further half swap for Row would create a game of Matching Pennies, *do-db*, located in Double Ties for both players. Even as a player becomes increasingly indifferent between various payoffs, the structure of the game remains cyclic. Thus, even when preferences are quite simple, with only two ranks, it is possible to have a cyclic structure with no equilibrium solution. Similarly, even with ties for high and low payoffs it is still possible to have a stag hunt game with two equilibria, Double Coordination, *do-do*.

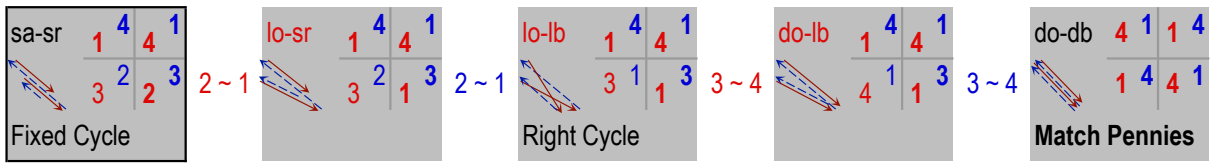

*Figure 14. Fixed Cycle turns into Right Cycle and then into Matching Pennies*

Remembering that in the strict topology the two cyclic games are linked by a hotspot shows how the cyclic games form a connected region within the expanded topology, including the games on the pathway between Fixed Sum and Matching Pennies. Thus, the game of Tragic Cycle, *A-B, or sa-sb,* lies at the heart of a refractory region, where swaps and half swaps mostly create other cyclic games. The adjoining zero-sum (fixed rank-sum) game and the other fixed sum games are the most difficult games to convert into win-win, requiring at least three swaps (Plate 2d). By contrast, the Harmony tile and its two adjoining Privileged tiles   and the pipes that connect them with the Benevolent, Generous and Second-best tiles form a more satisfying and stable neighborhood, where both players get at least their second-best result. The game of Harmony lies at the heart of a large, pleasant, region, including the High, Middle, Low, Double, Triple and First versions of Harmony and the three harmonious pipes that link harmonious, second-best, and biased/generous tiles.

While forming ties creates simpler games, the reverse process of breaking ties can be understood as showing the evolution of preferences and more complex joint incentive structures. This starts from the simplest zero game, where both players are completely indifferent and moves all the way to the complexity of Prisoner's Dilemma's symmetric Pareto-deficient equilibrium. Thus, the topology shows how more precisely differentiated preferences lead to more complex structures of interests, affecting the ease or difficulty of achieving cooperation. Apparent indifference between outcomes could be due to uncertainty, lack of information, or lack of attention. As more information becomes available, and more attention is devoted to analyzing alternatives, more detailed preferences may develop, shifting from a simpler to a more complex structure.

From an evolutionary point of view, the progression from simpler to more complex games shows the advantage of gaining increasing capabilities, from basic abilities to seek or avoid, to synchronizing behavior, anticipating a dominant strategy and the appropriate response, or agreeing on an alternative in a coordination game. Thus the complete set of 2x2 ordinal games illustrates the advantages of various capabilities, including basic sensitivities, anticipation, and analysis as well as more sophisticated capabilities for creating trust and coping with defection.

Where more precise distinctions than simple ranking are relevant, games with real value or interval-scale payoffs can also be normalized and mapped into the the expanded topology, as with Axelrod's version of Prisoner's Dilemma (Plate 2g). Thus, the ordinal games provide a grid of coordinates within the space of 2x2 games, and the visualization chart can be seen as displaying not only the ordinal games, but also the boxes of points containing normalized games with equivalent ordinal structures.[13] The lines between games are composed of games with ties. The probabilities of different games occurring at random would be mapped by the space within which ordinal games form discrete coordinates. Thus, the enhanced visualization maps not only the strict ordinal games, and the games with ties that form borderlines between them, but also all normalized 2x2 games and their possible payoff transformations.

## Discussion and Conclusions

Escaping from Prisoner's Dilemma is difficult. Robinson and Goforth (2005) describe the Prisoner's Dilemma family of Pareto-deficient games as a bad neighborhood. Prisoner's Dilemma's tragic neighbors yield similarly poor results. The nearest win-win game can be reached through swaps in the highest payoffs for both players, but turns out to be a precarious Stag Hunt, where distrust and risk avoidance could lead to results as bad as in Prisoner's Dilemma. Half swaps from Prisoner's Dilemma, equalizing the top two payoffs, create a borderline game, High Dilemma, which mixes the difficulties of an assurance game and Prisoner's Dilemma, since the win-win outcome would be a Nash Equilibrium, but following a weakly dominant strategy or minimizing risk leads to being trapped in the inferior outcome. The nearest stable solution to escaping Prisoner's Dilemma comes through swapping middle payoffs to create a game of Deadlock, where both have dominant strategies leading to their second-best outcome.

Other social dilemmas are more easily transformed into win-win games. A single swap in the highest payoffs for one or the other person turns a proper battle of the sexes into a proper coordination game. Swaps in highest payoffs for either person transform Samaritan's Dilemma into a harmonious Pure Aligned game. Swaps in high payoffs for both turn Chicken into a harmonious game of No Conflict.

This paper builds on Robinson and Goforth's (2005) topology of 2x2 ordinal games, and their (2007) article with Cargill expanding the topology to non-strict games, as well as earlier work by  Fraser and Kilgour (1986, and Kilgour and Fraser 1988) on nonstrict ordinal games and Rapoport and Guyer (1966, and Rapoport, Guyer, and Gordon 1976) on the taxonomy of 2x2 games, including borderline and degenerate games. Based on the analysis presented in the paper, it can be concluded that:

- The topology of 2x2 games maps potential transformations between games, including paths from social dilemmas to win-win games and paths of vulnerability by which win-win games could turn into games with worse outcomes.
- To the extent that payoffs occur randomly, Prisoners Dilemmas are likely to be relatively rare, only about 5% of all games, while a large majority of games (64%) allow both players to get their best or second-best outcome at equilibrium.
- There are 38 distinct 2x2 ordinal games, including the 12 strict symmetric games and 26 games resulting from half-swap transformations for making ties (or, conversely, for breaking ties starting from the Zero Game of complete indifference).
- The symmetric ordinal 2x2 games can be used as coordinates in a simple naming scheme to uniquely identify the complete set of 2x2 games, including a variety of games that may be interesting for future research.
- The enhanced visualization of the topology of 2x2 games, with borderlines formed by games with ties, maps not only ordinal games, but also normalized versions of all 2x2 games, and their potential transformations.
- The elementary models and relationships mapped by the topology of 2x2 games provide a tool for understanding the diversity of social situations, and the consequences of changes that switch the ranking of outcomes, including the results of deliberate institutional design.

On a more speculative note, one can wonder about numeric relationships in the expanded topology of 2x2 games with and without ties. As discussed by Robinson and Goforth (2005) the torus formed by the topology of strict 2x2 games has 37 holes. If the Zero game is not included, there are 37 symmetric ordinal 2x2 games. While formal analysis or proof is beyond the scope of this paper, the obvious conjecture is that the 37 holes and 37 games may not be numeric coincidences, and that the total number of 37 distinct symmetric 2x2 ordinal games (not counting the Zero game) is inherent to the structure of the topology.

The topology of 2x2 games offers a set of tools for mapping the transformations between games, and understanding the potential solutions and risks, that may result from side payments, reaching agreements, making rules, better information, refining preferences, improved monitoring and enforcement, technical innovation, and other changes. The unsatisfactory outcomes of Prisoner's Dilemmas and zero-sum games and the difficulty of remedying the through changing payoffs provides a justification for giving them disproportionate attention. Nevertheless, they should be understood within the diverse context of the majority of games that have better equilibrium outcomes and are more easily transformed into win-win games.

The Robinson-Goforth topology of two-person, two-strategy games provides an elegant and useful map of how games are related to each other, and the potential moves between games. It reveals a landscape of cooperation and conflict, elementary but meaningful "toy" models of how dual, and sometimes dueling, desires entwine. The topology can be applied to understand the structure and range of strategic interactions, and their transformations. The topology of 2x2 games extends to games with ties and normalized versions of games with interval-scale or real-value payoffs. The topology could also be extended to multiple players, moves, repetitions, and various payoff functions (trembling, fuzzy, superpositioned, etc.).

Ties make simpler games, including archetypal games of indifference, cooperation, conflict, cycles, and coordination, as well as simpler versions of Battle of the Sexes, Coordination, Second Best, and other games. The expanded topology of 2x2 games with ties locates a variety of interesting games that may be fruitful for further study such as those listed in Appendix A. The topology shows the elegant diversity of 2x2 games, illuminating the variety of possible models of strategic interaction beyond Prisoner's Dilemma and the paths for escaping from social dilemmas to create harmonious win-win games.

## APPENDIX - NOTES ON GAMES WITH TIES

This appendix provides some additional discussion of games with ties. The games themselves are shown in Plate 3 as part of an overview of the expanded topology of games with ties and in Plate 4 with numeric payoffs in the complete set of 2x2 games.

In the literature (including Robinson, Goforth and Cargill 2007), *non-strict* often seems to be used loosely to refer to the full set of 2x2 games, both with and without ties, which Rapoport, Guyer, and Gordon  (1976) call the *complete set*. Where precision is needed to distinguish those games with ties from those without ties, this paper uses the slightly more wordy, but more accurate terminology of *games with ties* to refer to 2x2 games with at least one pair of ties.

While relatively little attention has been paid to asymmetric strict ordinal games (aside from constant-sum games, also known as fixed-sum), nonstrict asymmetric games have received even less study. The list below identifies some potentially interesting games with ties, symmetric and asymmetric, from this study and a few other sources. RGG numbers refer to the table of 2x2 games at the end of Rapoport, Guyer, and Gordon's (1976) book, *The 2x2 Game.* F&K numbers refer to the key numbers in the pioneering articles by Fraser and Kilgour 1986 and Kilgour and Fraser 1988 on 2x2 games with ties.

**Additional Phenomena in 2x2 Games with Ties**

Compared to the strict 2x2 games, a variety of additional phenomena emerge in 2x2 games with ties. Several of these seem particularly interesting for modeling and experimental study of behavior. Fraser and Kilgour (1986) and Kilgour and Fraser (1988) also provide additional analysis of game characteristics and solution concepts for nonstrict 2x2 ordinal games, which might be usefully revisited in light of the topology of 2x2 games, though such analysis is beyond the scope of this paper:

- *Borderline* (Rapoport et al. 1976). Between two existing games, and sharing some of their characteristics, as for example in Low Dilemma between Chicken and Prisoner's Dilemma.
- *Degenerate* (Rapoport et al. 1976): One or both players are indifferent between the two choices, in other words, for each of the other player's strategies, the first player gets to same payoff. For one example, see the discussion in Simpson 2010: 42-43 (and Simpson 2011) and the discussion of his game of "Red Dress" below.
- *Weak Dominance* (Fraser and Kilgour 1986): A choice of strategy leads to a better result for one of the other player's strategies, while for their other strategy the first player is indifferent between the two outcomes. Often this is sufficient to create solution, but sometimes becomes problematic, especially if the result is pareto-inferior, as in the High Dilemma.
- *Limits to dominance-solvability:* The Strict games of Samson, *sb-sn*, and Delilah, *sr-sn*, already pose questions about the extent to which a player would rely on a myopic strategy of assuming the other player will always follow a dominant strategy.[14] For the game *mb-sn (*located between Samson, *sb-sn*, and Hegemony, *sc-sn),* as well as for its neighbor, *mb-ln,* it seems highly questionable whether a player would give up certainty of getting their second-best outcome and instead rely on the other player to choose a "weakly dominant" strategy since that leads the first player to get their worst outcome. Thus hegemony, based on one player being trapped by dominant strategy, arguably collapses into indeterminacy, a reductio ad absurdum for dominance solvability. Similar issues apply to Low Dilemma, and could be fruitful for experimental study.
- *Generosity and spite*: At no cost to themselves, one player can help or hurt the other's payoff.
- *Guaranteed*: One player can ensure that both win, or that she wins and the other loses.

- *Extreme Unfairness and Tragedy*: One player gets their worst payoff, while the other gets their best or second best. Extremely tragic games are pareto-inferior for one or both players.
- *Varieties of Pareto-optimality:* In the strict 2x2 games, games are Pareto-inferior for both, but in games with ties there may be alternatives that would make things better for one while leaving the other the same. This satisfies the usual economic definition of Pareto-optimality, but the implications for a game with strategic interaction between players may be more complex, especially if a shift to an apparently Pareto-optimal outcome creates a risk of doing worse. This has implications for the stability of solutions, since one player can change strategies with no immediate impact to her payoff. Where a distinction matters, Pareto-optimal can be used for the general meaning that there is no way to make anyone better off.  *Mutual Pareto-optimality* can be used for the situation where there is no superior alternative where both would gain.[15]
- *Interdeterminate Payoff Family*. In terms of payoff families, the degenerate games form a distinct family or subfamily of games, which can be combined with the cyclic games, and other games with no dominant strategy and no Nash Equilibrium into an extended family of indeterminate games. Most games with ties can be categorized using the same families as the strict games, though for games with ties the distinction between the unequal outcomes becomes less clear. In general, those with best and second best payoffs at equilibrium have been categorized as Biased, those with highly unequal outcomes as Unfair, and those with a weak dominance equilibrium at second-worst or worst payoff as in the Pd family.

**SYMMETRIC GAMES WITH TIES**

*Coordinates.* In the topology of the strict ordinal games, the symmetric games form a diagonal axis, and this also applies for the expanded topology that includes games with ties.[16]  There are a total of thirty-eight distinct symmetric ordinal games with ties, including the Zero Game of complete indifference. However, in forming games through half swaps, the high swaps generate duplicate games which are equivalents by interchanging rows and columns. Furthermore, payoffs from some of the duplicate games are needed to generate games derived from half-swaps, two cyclic games and two pairs of unfair games in the High Ties class and Matching Pennies and two unfair games in the Double Ties Class. A listing of the symmetric games, including the duplicates, is composed of forty-seven games arranged along a diagonal axis, as in Plates 3 and 5.

These games are located between the twelve ordinal games, formed by half-swaps that make sites. These relationships can be shown on the surface of a cube (Huartas-Rosero 2003, 2004, Goforth and Robinson 2011), with two equivalent halves, each composed of 12 game triangles, 18 edge games (low, middle, and high ties), and 7 vertex games (triple, double, and basic) as shown in Plate 4.

Having unique names for all of the symmetric games provides a relatively easy way to uniquely identify and locate the asymmetric games. While it would be possible to extend Robinson and Goforth's numeric indexing scheme to the games with ties, this becomes complicated and not particularly intuitive or easy to remember. By contrast, the symmetric games provide a convenient set of coordinates for locating games, and names and abbreviations make them easier to remember. Furthermore, naming based on the symmetric games easily accommodates different ways of displaying the topology, including keeping the preference classes distinct, as in Plate 3, or interlacing the strict games with low and middle ties, as in Plate 4.

Most of the symmetric names with ties can be given names based on the type of ties and related strict game, for example the various versions of Harmony. The Zero game is unique and distinctive enough that it is simpler to use a separate name, and this may also be the case for some other games, such as the Double Dilemma (Avatamsaka or Interdependence) Game. As with the strict games, the common name can be followed by coordinates, for example Volunteer's Dilemma, Middle Battle, *mb-mb*, formed by middle swaps from Battle, s*b-sb* or by middle swaps from its immediate neighbor, *sc-sc*. In the interest of space and simplicity, especially for tables and charts, the coordinates for symmetric games need not always be repeated, i.e. *mb* rather than *mb-mb.*

Zero
- ***ze* – Zero Game.** The "game" of utter indifference, with all ties and no preferences, is in a class by itself.

SINGLE
- **Basic Harmony**, *bh-bh*. The simplest version of harmony. where each player has only a single top-ranked outcome, and rates the other three equally. The simplest win-win game and archetype for all the games on Layer Three, the Win-win Layer
- **Basic Dilemma**, *bd-bd*. The game with the best payoffs diagonally opposite each other is the simplified, archetypal version of all the games on the Discord Layer (Layer One), and the conflict underlying Prisoner's Dilemma, Chicken, and the battles of the sexes.

DOUBLE TIES
- **Double Dilemma, Avatamsaka**, **Interdependence** *dd-dd. T*his interesting win-win game, lies "between" Stag Hunt, Prisoner's Dilemma, No Conflict and Chicken, formed by ties on the two highest and two lowest payoffs. It has no dominant strategy and no maximin solution. It is degenerate (Rapaport, Guyer, and Gordon 1976) in the sense that one or both players are indifferent between their choices. Aruka (2001, cited in Goforth and Robinson 2011) uses the name Avatamsaka Game, based on Buddhist scripture, where choice can lead to heaven or hell. This game exemplifies the ethical principle of the Golden Rule, "do unto others as you would have them do unto you," a principle which would allow converging on the Pareto-optimal outcome as a prominent focal point and Nash Equilibrium.  A "[d]egenerate no-conflict game" used to study competitive pressure (Rapoport et al. 1967 cited in Rapoport, Guyer and Gordon 1976: 31, 34, 226-227). RGG#79

LOW TIES

- **Low Dilemma,** *ld-ld*. The game between Prisoner's Dilemma and Chicken combines properties of both. It has been recognized in the literature, but does not seem to have an established name, so the name Low Dilemma is proposed here. Players have an incentive to defect from cooperation, leading to both getting their lowest payoff, but then have nothing to lose from switching to a cooperative strategy. Discussed by Rapoport, Guyer and Gordon (1976:31) as a "borderline" game between Prisoner's Dilemma and Chicken, but payoff matrix not presented. Fraser and Kilgour (1986:117-118) identify it as particularly interesting, saying that "preliminary analysis suggests it gives the players almost no way to cooperate, so that one must 'lose.' " F&K#216
- **Low Coordination,** *lo-lo*. Low ties make this a pure example of Sen's (1967) assurance problem where cooperation is best, but if the other does not cooperate then it is also best to not cooperate. The low ties version has no risk-avoiding (maximin) solution. This has been used as a "no-conflict" game to study a stochastic learning model (Suppes and Atkinson 1960 cited in Rapoport, Guyer and Gordon 1976:410-412). RGG#83
- **Low Battle of the Sexes,** *lb-lb*. The game between Hero and Leader is an idealized version of the Battle of the Sexes, with two equilibria, where one or the other player does better, and they are indifferent about the other two less-preferred outcomes.

MIDDLE TIES
- **Midlock**, *mk-mk. T*he game between Prisoner's Dilemma and Deadlock leads to both getting their second-best result, a unique equilibrium with no Pareto-superior alternative, and no dilemma, so it seems best to name it as a version of Deadlock. Note that this is a constant-sum game and the only symmetric game of pure opposition.
- **Middle Hunt, Rousseau's Hunt**, *mu-mu.* In Rousseau's classic story, a hunter must choose between catching a rabbit for sure or possibly sharing a stag if the other hunters cooperate . Since the impact on the other hunter is not discussed, this implicitly treats the first hunter as indifferent about the impact of their action on the other.
- **Middle Battle, Volunteer's Dilemma**, *mb-mb.* Each prefers that the other volunteer, for example calling 911, but if she does volunteer then she doesn't care if one or both contribute, while risking what both consider the worst result if neither volunteers. This can model a situation where one contribution is sufficient and a second contribution adds no benefits, so each would like the other to pay the costs. Identified by Fraser and Kilgour (1976: 117-118 #445) as a "destabilized version of chicken, since the cooperative outcome is much less likely to persist in these games than in Chicken." F&K#445.

HIGH TIES
- **High Dilemma**, *hd-hd.* Intermediate between Prisoner's Dilemma and the strict Stag Hunt, with two unequal Nash Equilibria, where weakly dominant strategies get trapped in a Pareto-deficient result, doing as badly as in Prisoner's Dilemma despite the presence of a win-win outcome that would also be a stable best response equilibrium. Located between the four games at the center of the topology chart, *Pd,*

*PdSh, ShPd, and Sh,* this is the epitome of a borderline game, located between the four layers, and between the Stag Hunts and Prisoner's Dilemmas. In terms of solution concepts, this is, arguably, perhaps the most complex game, combining weakly dominant strategies leading to a Pareto-deficient outcome with a second Pareto-superior equilibrium, (from which neither can unilaterally improve their payoff), in which achieving the win-win outcome faces coordination, assurance, and risk avoidance problems.

- **High Hero**, *he-he*. The game between either symmetric battle of the sexes, Hero or Leader, and a Pure Coordination Game (*so-sa* or *sao*) is pure, since it has the same values in each cell, and has the properties of a coordination game. It can be formed by combining payoffs from the High Coordination, *ho-ho* and its interchanged equivalent, *ha-ha*, as well as by creating high ties in Hero, *se* or Battle of the Sexes (Leader), *sb-sb*, and then switching columns to align the win-win payoffs on the left-right diagonal. In the chart, the version on the diagonal is emphasized, especially since its payoffs are also needed to generate the left and right cycle games.

## ASYMMETRIC GAMES WITH TIES

- **No Conflict-Zero**, *sn-ze*. One person has full power over the other player, but no power over her own outcome, so the game is "degenerate." (Rapoport, Guyer and Gordon 296-297, who cite research on this game by Swingle 1970). RGG#80
- **Double Safe**, *dd-dh.* One person's decision means both will win.
- **Double Unfair**, *dk-dd.* One person's decision means she must win and the other must lose.
- **Matching Pennies,** *do-db.* The simplest cyclic game and the simplest zero-sum game.
- **Red Dress,** (Double Avatamsaka-Low Prisoner's Chicken), *dd-ld***.** Discussed by Simpson (2010: 42-43, 46, 48) as an interesting nonstrict asymmetric game. A boy wants his girlfriend to wear a dress while he wears jeans, rather than him wearing a suit and her a dress, but he equally dislikes wearing a suit or jeans if she is in jeans. She would prefer he wear a suit, regardless of what she wears. For each of his strategies, she gets equal payoffs regardless of what he does, and so has no direct power to affect her own outcome, a "degenerate" game, but one in which her decision affects his payoffs.
- **Jekyll-Hyde** (Middle Deadlock-Invisible Hand) *mk-mh.* One kind, one cruel; the simpler Middle Ties version of the strict Type games (Robinson and Goforth 2005a, 2005b) where one person, for each of their strategies, is induced to raise payoffs for the other player, while the second player is induced to reduce payoffs for the first. One player's incentives always produce positive externalities and the other's incentives always produce negative externalities: one helps, one harms, and, unlike in the movies, the villain wins. A game for good guys to avoid, preferably through less tragic means than the conclusion to Robert Louis Stevenson's classic story. Located in the center of the *middle tile* formed by 2↔3 swaps between *sd-sn, sd-sh, sk-sn* and *sk-sh*, at the intersection of the quasi symmetric and sub-symmetric axes. Jekyll-Hyde can be transformed into a win-win game by either of two high swaps for each player. Three of the strict Jekyll-Hyde Type games can be converted into win-win by a single high swap for row. The unfair game of *sk-sn* is more refractory, located in the Left-Discord hotspot, with three unfair and one tragic game for neighbors along with the biased Jekyll-Hyde Type game *sk-sh*, from which two swap steps, either middle and row swaps for Row, or a middle swap for Column and high swap for row, would be needed to reach win-win.

- **Moral Hazard, Principal-Agent Problem**, (Rousseau's Hunt - Low Dilemma), *mu-ld.* A moral hazard or principal agent problem exists if cheating on an agreement, for example with an insurer or an employer, brings a higher payoff than fulfilling the agreement, typically if the insurer or employer is unable to adequately monitor behavior and enforce penalties for noncompliance. If the insurer or employer is indifferent between avoiding cheating and missing the opportunity to contract but prefers either of those to being cheated, and the other party is also indifferent between missing the chance to cheat or missing a chance to cooperate, and but likes either of those less than cooperating or cheating, then the structure combines the payoffs of Rousseau's Stag Hunt, *mu,* and Low Dilemma, *ld*. A solution is to swap the payoffs for cheating and cooperating, by increasing the penalties and risks associated with cheating or increasing the benefits of cooperating, thereby transforming it into a win-win Principal-Agent game, *mu-ln*, where the dominant strategy for each is to honor the contract.

- If the insurer or employer moves first, in deciding to offer a contract, then if Row expects defecting to pay off better for Column, Row would take the safer, maximin, option of never offering a contract, as is made obvious by the extensive form game tree. However, if they expect the contract to be honored, thinking they are playing a win-win game, they could enter the contract, but then face changed conditions that shift the payoff structure (or reveal that they misjudged the situation). Thus, the topology shows the risks if conditions change enough to swap the highest payoffs, indicating the potential instability and vulnerability to post-contractual opportunism. If both parties have entered a contract where each would be better off defecting, for example breaking the promise to pay insurance, but the result from both cheating is worse than both cooperating, then this breaks the ties in Moral Hazard, to produce the familiar payoff structure of Prisoner's Dilemma. Breaking the ties the other way, so the stronger party would be tempted to cheat, while the other, like a serf or criminal, might be inclined to act cooperatively even with someone known to cheat, but would then defect once the other starts to cooperate, moves in the opposite direction to create a cyclic game, Inferior Cycle, *a-c*, with no stable equilibrium in pure strategies.

- **Middle Compromise – Rousseau's Hunt**, *mm-mu.* Degenerate game of complete opposition. Used to study a stochastic learning model (Suppes and Atkinson 1960, cited in Rapoport, Guyer and Gordon 1976:406-409. RGG#81

- **Middle Compromise-Peace**, *mm-sp.* Partial conflict game, similar to RGG#15 (Subsymetric Benevolence, ReDe). Used to study a stochastic learning model (Suppes and Atkinson 1960 cited in Rapoport, Guyer and Gordon 1976:410-411). RGG#82

- **Middle Compromise-Prisoner's Dilemma**, *mm-sd.* Identified by Fraser and Kilgour (1976:117-118) as one of ten nonstrict ordinal games with properties similar to

Prisoner's Dilemma. F&K#507
- **Destabilized Chicken** (Volunteer's Dilemma-Chicken), *mb-sn*. Identified by Fraser and Kilgour (1976: 117-118), along with Volunteer's Dilemma, as a "destabilized" version of chicken, "since the cooperative outcome is much less likely to persist in these games than in Chicken." F&K#533

# Notes

The names for symmetric games used here follow common usage , and mostly match those in Robinson and Goforth's (2004) "Periodic Table of 2x2 Ordinal Games," with some modifications:

- Deadlock and Compromise are used as more distinctive names for the symmetric second-best games adjoining Prisoner's Dilemma and Hero, rather than Anti-Prisoner's Dilemma (#155) and Anti-Chicken (#166). Note that Binmore (2005, 2007) shows the payoff matrix for Compromise, with the name Prisoner's Delight, but also uses the term more generally to refer to games where dominant strategies lead to both players getting next-best or best payoffs.

- Battle of the Sexes is used for the game Rapoport (1967) named Leader (#133) next to Chicken,• The symmetric two-equilibrium game next to Stag Hunt (#333) is a more severe example of Sen's (1967) assurance problem (dropping from best to worst if the other defects) providing a way to disambiguate names for the two coordination games., so the remaining coordination game (#344), where the win-win outcome is also the maximin solution, can be called Coordination.

- The unnamed symmetric game (#355) on the same tile as Harmony has mixed interests, since each player's shift from lowest to second-lowest payoff reduces the payoff for the other player. Despite this initial divergence in interests, dominant strategies still converge on the best result for both. This movement from discord to concord suggest the name Peace, emerging from initial differences into agreement.

[5] This assumes that games that swap columns, rows, or both can be treated as equivalent, otherwise the number would be even larger, with 576 variants. If necessary, these "interchanged" variants can be labelled according to the quadrants they occupy, analogous to Cartesian coordinates, for example as northeast, southeast, southwest, and northwest variants.

[6] In terms of conventional game theory, this can be thought of as adding another strategy, which changes a 2x2 game into a 3x3 game. If the new strategy dominates one of the existing strategies, then the "new" game simplifies into a different 2x2 game, in other words one ordinal game has been transformed into another.

[7] The proper cyclic tiles double-link Layers Two and Four. The tragic-unfair tiles double-link Layer One with Layers Two and Four, while the altruistic and aligned tiles double-link Layers Three and Four. *Pipes,* such as the Pd Pipe, link four tiles on four layers. Plate 2d shows how hotspots and pipes link equivalently-located tiles across layers.

High swaps can be seen as tunneling through the third dimension (or a space warp) to link games on different layers (Robinson and Goforth 2005). As another way to think about it, high swaps can also be seen as roughly analogous to a knight's move in chess, with a major component that switches layers, and a minor component that switches row or column. The destination layer depends on the final alignment of 4s, and so depends on where the swapped 3 was in the original game, relative to the other player's 4.

[8] Note that this high swap transformation from Volunteer's Dilemma to a win-win game occurs through the hotspot linking the battles of the sexes and coordination games. The two columns are interchanged to keep Row's 4 on the right.

[9] In the same way that swapping the two lowest payoffs creates a tile of four games, swapping middle payoffs also links four games. Robinson and Goforth (2005) only discuss tiles formed by swaps in the lowest payoffs, but the concept can be extended to middle and high swaps. Thus, the Jekyll-Hyde Game, *mk-mh* lies in the center of what can be called a *middle tile* formed by the four strict Jekyll-Hyde Type games (*sd-sn, sd-sh, sk-sn, sk-sh*). Similarly, High Dilemma, *hd-hd*, lies in the center of a *high tile* joining Prisoner's Dilemma and Stag Hunt.

[10] This table leaves out duplicate games that are equivalent by interchanging rows and columns. However, interchanged versions of symmetric games are needed in the Double Tie and High Tie classes to generate the games outside of Layer Three, such as Matching Pennies.

Where possible, duplicates have been pruned to create a compact, contiguous, rectangular (toroidal) matrix,with as many games as possible in the lower left, win-win, layer. However, the asymmetry of pipes and hotspots means that the high ties and double ties classes and their combinations cannot be restricted to a compact, contiguous matrix.

The strict game algorithm of "Row's 4 right, column's 4up" is insufficient to uniquely arrange nonstrict games within a class. To find the relevant game in the table it may be necessary to interchange rows or columns. Identifying the Nash Equilibrium(s) (or lack thereof) may also facilitate locating the game in the table.

[11] Kilgour and Fraser (1988:109) state that there are thirty symmetric ordinal games. This

would be the total for the twelve strict games and eighteen games with high, middle, or low ties, excluding the games with only two ranks (triple, double, and basic) and the null game. However, they discuss the Null Game, their Game #1 (p. 111), as one of the symmetric games, and count only 10 strict ordinal symmetric games (Figure 4. p. 112) so it is not clear where the discrepancy comes from. Note that Fraser and Kilgour (1986, and K&F 1988) analyze the characteristics of different nonstrict 2x2 games, and the applicability of various solution concepts.

[12] Matching Pennies can be formed by combining payoffs from the double ties coordination game,, *do-do*, and the double ties version of Hero and Battle, db-db equivalent to *do-do* with rows and columns interchanged. For Matching Pennies, the display used here generates two versions, equivalent by interchanging rows and columns. If a choice is necessary, the right-hand "counterclockwise" version could be designated as typical.

In the three-dimensional space of the topology's 37-hole torus, Matching Pennies is in the center of the cyclic hotspot that links Layers Two and Four.

[13] Robinson and Goforth explain that each ordinal game represents "an equivalence class" of other games whose payoffs have the same ordinal ranks. They present the formula for converting 1,2,3,4 ordinal ranks to fit the payoff scale in the version of Prisoner's Dilemma used by Axelrod. However, they do not discuss transformations in the other direction from cardinal (or interval) values, which provide a way to locate normalized versions of any 2x2 game within the topology.

[14] For a thorough discussion of *sb-sn,* see Brams (1994) discussion in terms of the Biblical Story of Samson and Delilah and Brams' Theory of Moves. The name Deliah for *se-sn* is based on its also having an alternative 3,3 outcome, but does not necessarily fit in the logic of Bram's narrative.

[15] The apparently standard term, "weak Pareto-optimality" refers to the situation where all would gain, and strong Pareto-optimality to subset of "strict' Pareto-optimality where at least one would gain. However, since this usage is potentially confusing and counterintuitive, the term mutually Perato-optimal is used here for the case where both would gain. See the Wikipedia article on Pareto-optimality, including the discussion page for further discussion of this terminology.

[16] In the three dimensional topology, this axis is part of one of two Villarceau circles (Robinson and Goforth 2005), which also forms the subsymmetric axis on the left and right layers. The other Villarceau circle is the axis of quasi-symmetry, delineating the fixed-sum games. The pure cooperation, pure opposition, and Jekyll-Hyde Type games are clustered at the intersections of these two axes.

Movement between simple and more complex games, between the null game and the surface of strict games, seems to represent movement in another two "dimensions," one for row and one for column, which converge at the null game.

Since a 37-hole toroid maps the topology on a three-dimensional object, half-swaps represent a movement not in a spatial dimension, but still seem orthogonal to the "plane" of the Villarceau circles. The diagrams in Plate 3b and 3d provide a way to visualize this movement, but further formal analysis of the expanded topology or better visualizations could be useful in understanding this.

# References

A bibliography for the topology of 2x2 games is available at www.bryanbruns.com/2x2chart

# 2x2 Games

**Payoff swaps change one game into another**

Adjacent ordinal games are neighbors by payoff swaps

1↔2 swaps form tiles of 4 games

2↔3 swaps join tiles into 4 layers

3↔4 swaps link layers

Layers differ by alignment of 4s

Each layer is a torus, table is a torus

Layers scrolled to center Prisoner's Dilemma

Symmetric  *Pure Cooperation*  *Pure Conflict*  Fixed Sum  *Jekyll-Hyde Type*

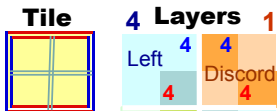

**Game**  Column payoffs  **Graph**
Row payoffs
Nash Equilibrium
Maximin
**Prisoner's Dilemma**  Pareto-deficient
Pareto-optimal outcomes in **bold**
Incentives & externalities
Mixed Strategy: (down, left) 3/4, 1/2
Payoffs 2.5, 2.5 iff ratio or real values

**Tile**  **Layers**
Left  Discord
Win-win  Right

| Families | Harmonious |
|---|---|
| 1. Win-win 4,4 | Stag Hunt |
| 2. Biased 4,3 | Battle |
| Self-serving | Samaritan |
| **3. Second Best** | 3,3 |
| 4. Unfair 4,2 | Winner  Loser |
| 5. Dilemma | Alibi 3,2 |
| Prisoner's D. 3,2 | Tragic 3,2 |
| 6. Cyclic (2.5,2.5) | |

**4**  Hotspots in serif  **Chart**  *A map of relationships between the 2x2 games*  **1**

[Grid of 2x2 game tiles — see chart image]

**3** To find a game: Make ordinal 1<2<3<4. Put column with Row's 4 right, row with Column's 4 up. Find layer by alignment of 4s; then intersection of Row & Column payoffs.

*see Robinson & Goforth 2005 The Topology of the 2x2 Games: A New Periodic*  www.cs.laurentian.ca/dgoforth/home.html  © CC-BY-SA ©2012.06.08  www.bryanbruns.com/2x2chart  **2**

# Structures in the Topology of 2x2 Games

**a. Symmetric games form a diagonal axis**

Chicken / Hawk-Dove/Snowdrift
Battle of the Sexes/Leader
Hero
Compromise  Anti-Ch
Deadlock  Anti-Pd
PRISONER'S D DILEMMA
Stag Hunt
Assurance
Coordination
Peace
Harmony
No Conflict, Concord

Payoffs

Symmetric   Quasi-symmetric   Sub-symmetric

**b. Twelve payoff patterns generate 144 games**

see Dragonica's Game Theory tonts: www.hf-fr-dragonicgames.com

Reflections around axis switch row & column positions
4 games per tile, 36 games and 9 tiles per layer
66 asymmetric pairs: 66 + 12 = 78 "unique" games

**c. Dominant strategies lead to equilibria**

Dominant strategies     Nash Equilibria

| D₁ Column | D₀ None |
| D₂ Both | D₁ Row |

1 0/2 (for ordinal
1 1 payoffs)

**d. High swaps (3↔4) link layers**

connecting equivalently-located tiles
6 Hotspots double-link pairs of tiles
13 Battle-Coordination Hotspot
6 Pipes link stacks of 4 tiles on 4 layers
NE Pd Pipe (Pd scrolled to NE corner to unify tiles)

High swaps switch up/down or left/right in tile

North  North
West  24  East  West  13  East
SW  South  14  SW  South  14
North  NE  North  NE
West  13  East  West  24  East
SW  South  23  SW  South  23

**e. Distance to win-win**

Swaps to reach win-win  1= single 3↔4 swap
2 or 3 step paths may include 2↔3 and 1↔2 swaps
# Bold = paths for both. Pareto-efficient paths,
each step results in same or better-ranked outcome
Zero-sum games are hardest to remedy

**f. Interests opposed, aligned, or mixed**

Externalities and inducement correspondences

Jekyll-Hyde Type ++  Pure Cooperation
Pure Conflict − −  Jekyll-Hyde Type
Fixed Rank-sum − ±  ± ++
(Zero Sum) see Schelling 1963 The Strategy of Conflict
Greenberg 1990; Robinson & Goforth 2005

**g. Games with ties form borderlines**

Games with ties (non-strict) lie between strict ordinal
games, at grid intersections (graph nodes/vertices)
Strict   Half-swaps: 2↔1   2↔3   3↔4
see Robinson, Goforth and Cargill 2007
The topology also maps all normalized 2x2 games

mu  Rousseau's Hunt
mb  Mid Volunteer's D
Axelrod's Pd

**h. Rapoport & Guyer Taxonomy**

|  | N | H | P | O | A | U | D | K | M | R | B | C |  |
|---|---|---|---|---|---|---|---|---|---|---|---|---|---|
| C | 55 | 50 | 49 | 70 | 78 | 72 | 39 | 35 | 35 | 65 | 67 | 66 | C |
| B | 56 | 52 | 51 | 74 | 76 | 71 | 37 | 33 | 32 | 64 | 68 | 63 | B |
| R | 44 | 41 | 40 | 75 | 77 | 73 | 31 | 34 | 36 | 54 | 55 | 51 | R |
| O | 18 | 16 | 15 | 53 | 42 | 43 | 10 |  | 30 |  | 17 | 14 | O |
| A | 23 | 21 | 20 | 57 | 47 | 48 | 12 |  | 10 | 38 | 46 | 37 | A |
| U | 26 | 22 | 23 | 58 | 62 | 61 | 48 | 46 |  | 61 | 71 | 72 | U |
| D | 27 | 24 | 25 | 59 | 63 | 62 | 47 | 45 | 72 | 60 | 64 | 65 | D |
| K | 30 | 28 | 29 | 60 |  | 67 | 57 | 53 | 74 | 75 | 76 | 78 | K |
| M | 2 | 4 | 3 | 29 | 25 | 24 | 20 | 13 | 13 | 40 | 42 | 59 | M |
| R | 8 | 6 | 5 | 25 | 23 | 21 | 15 | 12 | 10 | 36 | 47 | 63 | R |
| B | 7 | 3 | 8 | 30 | 27 | 26 | 17 | 14 | 11 | 54 | 56 | 55 | B |

Stable  Weakly Stable  Unstable

No Conflict  E=Natural outcome is Equilibrium
Mixed M EP D₂0 D₁0  D₂f D₁f D₂c 210 Dominant strategies
Motive M Ep D₂o pareto-deficient D₁f D₁tf no pressures
M e no e₂o e₁o competitive pressure
M E dominance D₀ No equilibria e₂o e₁c threat-vulnerable
Complete Z E D₂o D₁o e eD₂c force-vulnerable

Opposition  see Rapoport et al.1976 The 2x2 Game; Robinson & Goforth 2003

**i. Brams Typology and Game Numbers**

Number of non-myopic
equilibria (NMEs)
3 2 1
All Nash equilibria are
NMEs except asymmetric
pareto-deficient: 27,28,48

Not cyclic   Strongly cyclic
No conflict   Moderately cyclic
Weakly cyclic

see Brams 1994 Theory of Moves;
Difficult  see Brams & Kilgour 2008

© CC BY-SA  © 2012.06.08  BryanBruns@BryanBruns.com

# 3. Making Ties and Breaking Ties: The Complete Topology of 2x2 Ordinal Games

## a. Preference Classes: Type of Ties — Number of games (including reflections)

| Type of Ties | | | A | B | C | E | G | F | D | H |
|---|---|---|---|---|---|---|---|---|---|---|
| Strict | STRICT 1,2,3,4 H | 6 | 24 | 36 | 24 | 72 | 72 | 72 | **144** |
| Low Tie | 1,1,3,4 D | 1 | 12 | 18 | 12 | 36 | 36 | **36** | 72 |
| Middle | EDGE 1,3,3,4 F | 1 | 12 | 18 | 12 | 36 | **36** | 36 | 72 |
| High Tie | 1,2,4,4 G | 3 | 12 | 18 | 12 | **36** | 36 | 36 | 72 |
| Triple | 1,4,4,4 E | 1 | 4 | 6 | **4** | 12 | 12 | 12 | 24 |
| Double VERTEX | 1,1,4,4 C | 1 | 6 | **12** | 6 | 18 | 18 | 18 | 36 |
| Basic | 1,1,1,4 B | 1 | **4** | 6 | 4 | 12 | 12 | 12 | 24 |
| Null | ORIGIN 0,0,0,0 A | **1** | 1 | 3 | 1 | 3 | 3 | 3 | 6 |

**Total 1,413**

### b. Half-swaps make or break ties

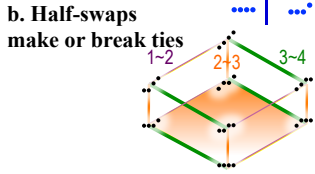

Figures a & b adapted from Robinson, Goforth and Cargill 2007, and see Fraser and Kilgour 1986, K&F 1988

## c. The 38 symmetric 2x2 ordinal games form a diagonal axis of symmetry

## d. Swaps & half-swaps link 2x2 Ordinal Games

## e. To find a game:
Make ordinal: Lowest = 1  Highest = 4  Middle ties = 3. Find class by type of ties, for each player. Put column with Row's 4 right, row with Column's 4 up.
Find Layer by alignment of 4s, then intersection of Row and Column payoffs. For high, double, and triple ties, interchange rows and columns if necessary.

# Transforming Symmetric 2x2 Games

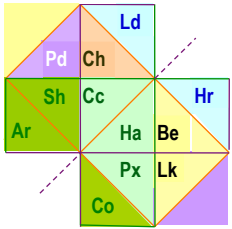

**a. Payoff swaps link strict symmetric games**

The twelve symmetric strict ordinal games are linked by payoff swaps

**High** 3↔4   **Middle** 2↔3   **Low** 1↔2

Normalized 2x2 games lie within the triangle for each strict game

Games with ties form borderlines between strict games

Ordinal games with low, middle, and high ties are at edge midpoints

Double tie games are at junction of four games

Basic and triple tie games are at vertices of six games

Reconciliation Line   $2c = s + t$
mk-mk,ld-ld,C,th-th,M,lk-lk,mk-mk

- - - Taking turns pays better
— Cooperative outcome pays better

Goforth and Robinson 2011 Effective Choice in All the Symmetric 2x2 Games

*from discord to harmony*

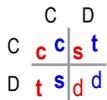

|   | C | D |
|---|---|---|
| C | **c** c | **s** t |
| D | **t** s | **d** d |

**Games on lower right edge interchange c & d, s & t:**
*do-do, bd-bd,tk-tk, do-do*

**c. Games with ties lie between strict games, on edges and vertices**

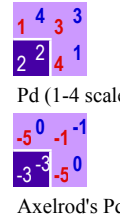 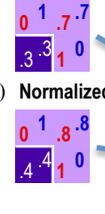
Pd (1-4 scale)   Normalized

Axelrod's Pd   Prisoner's Dilemmas

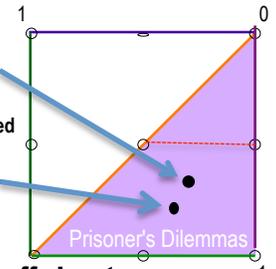

**b. Normalizing payoffs locates games**

**d. 38 Games in 8 Preference Orders**

| Null | Strict |
|------|--------|
| A 1,0,0,0 | H 12 1,2,3,4 |

7 Vertex 0 / 1    18 Edge 0 / .5 / 1

Basic  B 2  0,0,0,1
Low Ties  D 6  0,0,.5,1

Double  C 3  0,0,1,1
Middle Ties  F 6  0,.5,.5,1

Triple  E 2  1,1,1,1
High Ties  G 6  0,.5,1,1

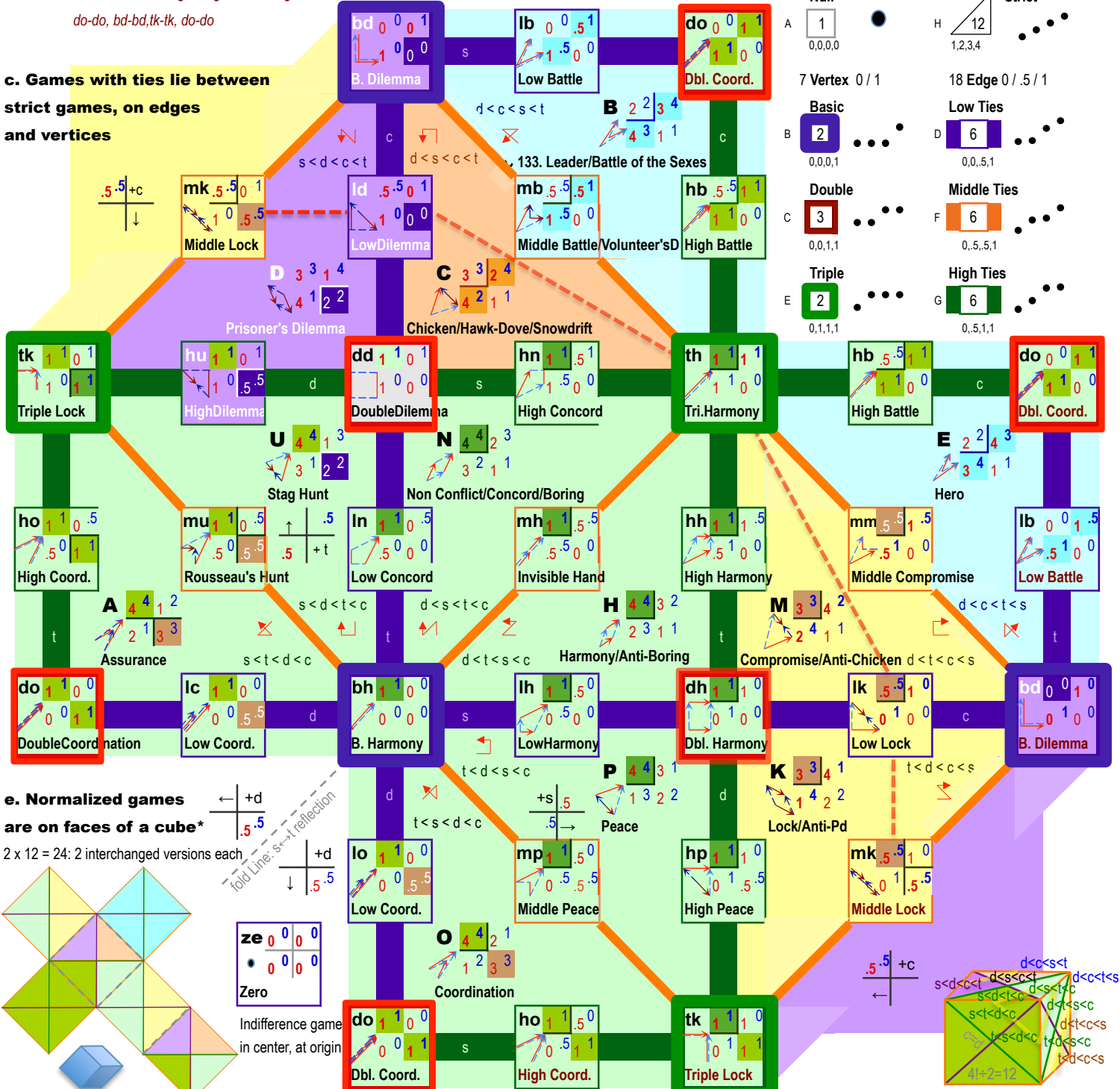

**e. Normalized games are on faces of a cube\***

2 x 12 = 24: 2 interchanged versions each

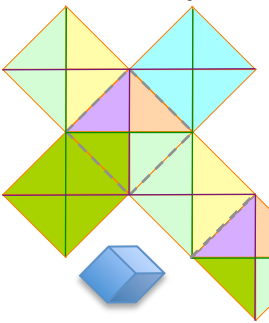

Indifference game in center, at origin

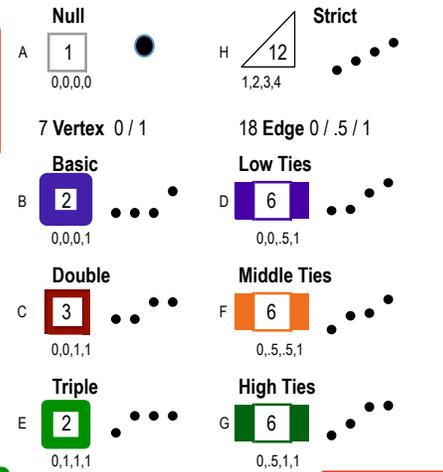

\* Disdyakis Cube, aka. Tetrakis hexahedron  v=14  f=24  e=36  74/2=37 games + Indifference Game = 38 symmetric 2x2 games

# The Symmetric 2x2 Ordinal Games

| | Index | | | | S | Names | ⋰ | | | RGG# | Brams# | F&K |
|---|---|---|---|---|---|---|---|---|---|---|---|---|
| **FACE** | | **HH** | | | | **Strict** | | **1234** | | | | |
| D | 1 | 111 | | | | Prisoner's Dilemma | | | | 12 | 32 | |
| C | 2 | 122 | | | | Chicken, Hawk-Dove, Snowdrift | | | | 66 | 57 | |
| B | 3 | 133 | | | | Battle of the Sexes, Leader, Bach or Stravinsky | | | | 68 | 54 | |
| R | 4 | 144 | | | | Hero, Battle of the Sexes, Bach or Stravinsky | | | | 69 | 55 | |
| M | 5 | 155 | | | | Compromise, Anti-Chicken, Prisoner's Delight | | | | 7 | 7 | |
| K | 6 | 166 | | | | Lock, Deadlock, Anti-Prisoner's Dilemma | | | | 9 | 9 | |
| N | 7 | 311 | | | | No Conflict, Concord | | | | 5 | | |
| U | 8 | 322 | | | | Stag Hunt | | | | 61 | | |
| A | 9 | 333 | | | | Assurance, Coordination | | | | 63 | | |
| O | 10 | 344 | | | | Coordination, Anti-Coordination | | | | 60 | | |
| P | 11 | 355 | | | | Peace, Resolution | | | | 5 | | |
| H | 12 | 366 | | | | Harmony | | | | 3 | | |
| | | | | | | | | | | | | |
| **EDGE** | | **DD** | | | | **L** | **Low Ties** | ⋰ | **1134** | | | | |
| ld | 13 | 111 | | | | Low Dilemma | | | | | | 216 |
| lb | 14 | 122 | | | | Low Battle, Battle of the Sexes | | | | | | |
| lk | 15 | 133 | | | | Low Lock | | | | | | |
| ln | 16 | 311 | | | | Low Concord, Low No Conflict | | | | | | |
| lo | 17 | 322 | | | | Low Coordination, Coordination | | | | 83 | | |
| lh | 18 | 333 | | | | Low Harmony | | | | | | |
| | | **FF** | | | | **M** | **Middle Ties** | ⋰ | **1334** | | | | |
| mb | 19 | 111 | | | | Middle Battle, Volunteer's Dilemma, Middle Leader | | | | | | 445 |
| mm | 20 | 122 | | | | Middle Compromise | | | | | | |
| mk | 21 | 133 | | | | Midlock, Middle Deadlock | | | | | | |
| mu | 22 | 311 | | | | Middle Hunt, Rousseau's Hunt | | | | | | |
| mp | 23 | 322 | | | | Middle Peace | | | | | | |
| mh | 24 | 333 | | | | Middle Harmony, Invisible Hand | | | | | | |
| | | **GG** | | | | **H** | **High Ties** | ⋰ | **1244** | | | | |
| *hd* | *25* | *111* | | | | *High Dilemma = hu, High Hunt* | | | | | | |
| *hc* | *26* | *122* | | | | *High Chcken = hn, High Concord* | | | | | | |
| hb | 27 | 133 | | | | High Batle, High Leader | | | | | | |
| *hr* | *28* | *144* | | | | *High Hero = hb, High Battle* | | | | | | |
| *hm* | *29* | *155* | | | | *High Compromise = hh, High Harmony* | | | | | | |
| *hk* | *30* | *166* | | | | *High Lock = hp, High Peace* | | | | | | |
| hn | 31 | 311 | | | | High Concord, High No Conflict | | | | | | |
| hu | 32 | 322 | | | | High Dilemma, High Hunt\ | | | | | | |
| *ha* | *33* | *333* | | | | *High Assurance = ho, High Coordination* | | | | | | |
| ho | 34 | 344 | | | | High Coordination | | | | | | |
| hp | 35 | 355 | | | | High Peace | | | | | | |
| hh | 36 | 366 | | | | High Harmony | | | | | | |
| | | | | | | | | | | | | |
| **VERTEX** | | **EE** | | | | **T** | **Triple Ties** | •••• | **1444** | | | | |
| tk | 37 | 100 | | | | Triple Lock | | | | | | |
| th | 38 | 300 | | | | Triple Harmony | | | | | | |
| | | **CC** | | | | **D** | **Double Ties** | ••ᵒᵒ | **1144** | | | | |
| *dd* | *39* | *111* | | | | *Double Dilemma = du, Interdependence, Avatamsaka* | | | | 79 | | |
| *db* | *40* | *122* | | | | *Double Battle = do, Double Coordination* | | | | | | |
| *dk* | *41* | *133* | | | | *Double Lock = dh, Double Harmony* | | | | | | |
| du | 42 | 311 | | | | Double Dilemma, Avatamsaka, Interdependence, Double Hunt | | | | 79 | | |
| do | 43 | 322 | | | | Double Coordination | | | | | | |
| dh | 44 | 333 | | | | Double Harmony | | | | | | |
| | | **BB** | | | | **B** | **Basic** | •••• | **1114** | | | | |
| bd | 45 | 100 | | | | Basic Discord | | | | | | |
| bh | 46 | 300 | | | | Basic Harmony | | | | | | |
| | | | | | | | | | | | | |
| **ORIGIN** | | **AA** | | | | **Z** | **Zero** | •••• | **0000** | | | | |
| ze | 47 | 000 | | | | Zero, Indifference, Null | | | | | | 001 |

# 2x2 Ordinal Games: The Complete Set

## Archetypal Games

### 4. High Ties (Pipe & Hotspot)

### 2. Double Ties

### 3. Triple Ties (Dislike)

### 1. Basic

### 0. Zero

# 2x2 Ordinal Games: The Complete Set

**Checkerboard interlace of strict with low and middle ties**

ln  N  mh  H  lh  P  mp  O  lo  A  lu  U  D  mk  K  lk  M  mm  R  lb  B  mb  C  ld

# 2x2 Ordinal Games: The Complete Set

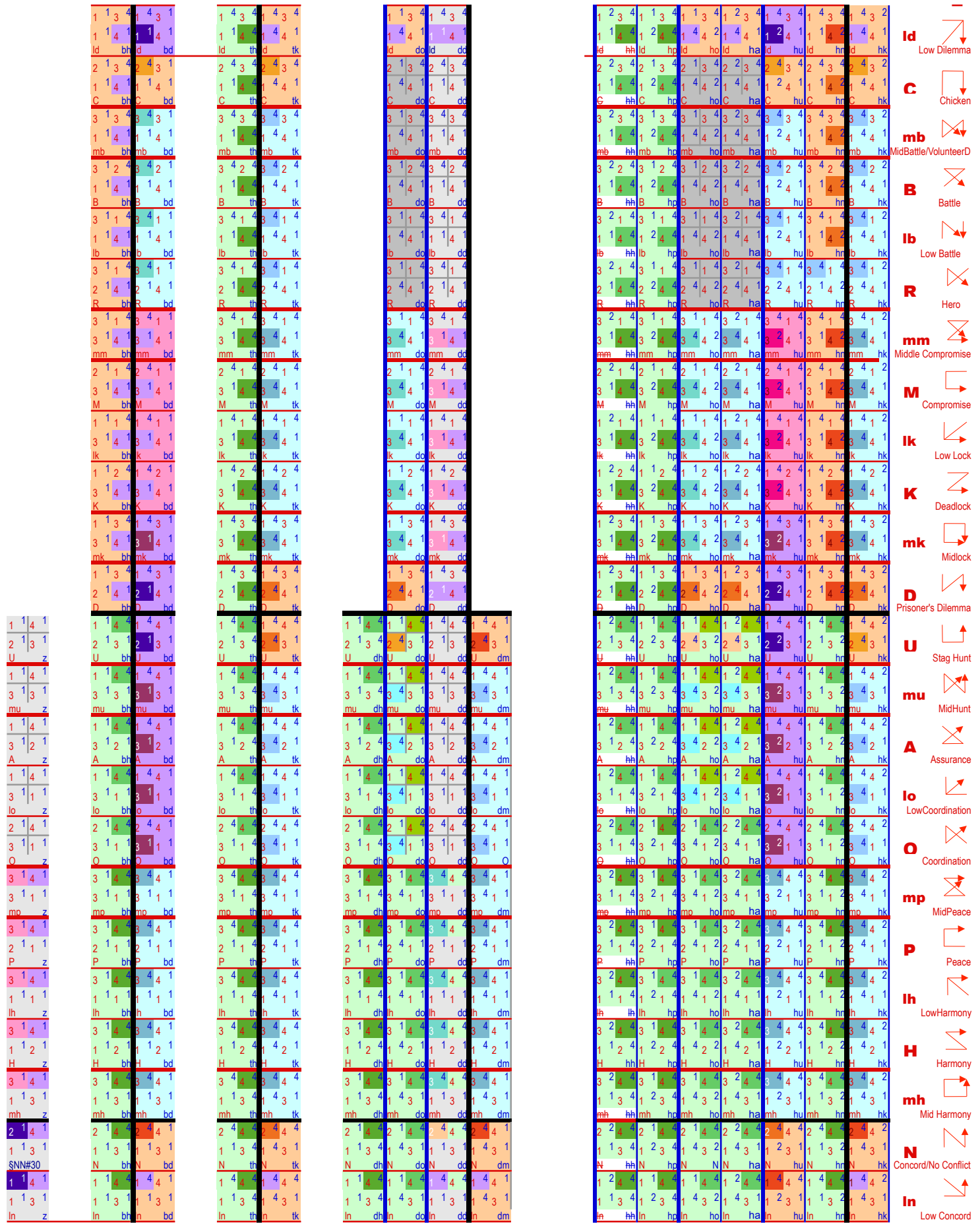

**Checkerboard Interlace (strict, low, and middle) with Archetypal**

ze     bh   bd     th   tk     dh   do   dd   dm   db     hh   hp   ho   ha   hh   hn   hk   hb

# 2x2 Ordinal Games: The Complete Set

## Archetypal with checkerboard interlace (strict, low, and middle)